\documentclass[aps,prb,twocolumn,superscriptaddress]{revtex4-1}

\usepackage{amsmath}
\usepackage{graphicx}
\usepackage{color}
\usepackage{combelow}

\begin{document}

\title{Quantum interference assisted spin filtering in graphene nanoflakes}

\author{A.~Valli} 
\author{A.~Amaricci}
\author{V.~Brosco}
\author{M.~Capone}
\affiliation{Scuola Internazionale Superiore di Studi Avanzati (SISSA)\\ 
CNR-IOM DEMOCRITOS, Istituto Officina dei Materiali, Consiglio Nazionale delle Ricerche, \\
Via Bonomea 265, 34136 Trieste, Italy}

\begin{abstract}
We demonstrate that hexagonal graphene nanoflakes with zigzag edges display quantum interference (QI) patterns analogous to benzene molecular junctions. 
In contrast with graphene sheets, these nanoflakes also host magnetism. 
The cooperative effect of QI and magnetism enables 
spin-dependent quantum interference effects 
that result in a nearly complete spin polarization of the current, 
and holds a huge potential for spintronic applications. 
We understand the origin of QI in terms of symmetry arguments, 
which show the robustness and generality of the effect. 
This also allows us to devise a concrete protocol 
for the electrostatic control of the spin polarization of the current 
by breaking the sublattice symmetry of graphene, 
by deposition on hexagonal boron nitride, 
paving the way to switchable spin-filters. 
Such a system benefits of all the extraordinary conduction 
properties of graphene, and at the same time, 
it does not require any external magnetic field 
to select the spin polarization, 
as magnetism emerges spontaneously at the edges of the nanoflake. 
\end{abstract}

\maketitle

The observation of interference patterns in the scattering and transport of particles
is a cornerstone of quantum mechanics which directly reflects the wave nature of matter. 
The very existence of quantum interference (QI) in a system demonstrates that its
transport properties are controlled by the laws of quantum mechanics.
In the last few years, evidence has grown that QI survives in molecular junctions 
like benzene,\cite{cardamoneNL6,staffordNtech18,solomonJCP129,stadlerNT14,stadlerNT15} 
and other $\pi$-conjugated systems\cite{solomonNC2,sahaPRL105,markussenNL10,markussenPCCP13,markussenPRB89}, 
up to relatively large molecules,\cite{gerlichNatComm2,richertNatComm8} 
thus showing that inherently quantum mechanical effects 
can be observed and exploited in the macroscopic world, 
which is very promising for the realization of electronic devices. 

In particular, zeroes of the conductance, or QI antiresonances can be exploited
to design QI devices as single-electron transistors\cite{cardamoneNL6,staffordNtech18}.
Predicting and controlling\cite{keNL8,frisendaNatChem8,richertNatComm8} these
antiresonances is therefore of paramount importance in light of applications.
In this respect, a set of graphical rules\cite{stadlerNT14,stadlerNT15,markussenNL10,markussenPCCP13,markussenPRB89} 
and their "diagrammatic" generalization\cite{pedersenJPCC119} 
represent important steps forward in the theoretical 
understanding\cite{xiaNL14,stadlerNT14,stadlerNL15,strangeNL15,zhaoJPC146} 
and classification\cite{pedersenPRB90} of QI features in molecular junctions. 

Recently, QI effects were also observed in graphene, in different contexts, 
including junctions,\cite{munarrizNanotech22,nozakiJPCL6} nanostructures\cite{nitaPSSRRL8,gengJACS137,sangtarashNanoscale8} 
and nanoconstrictions.\cite{gehringNL16} suggesting that what we learned for molecules is also relevant for complex nanostructures.

In this work we add a major twist to this scenario thanks to a  clear theoretical evidence for QI antiresonances for graphene nanoflakes with zigzag edges (ZGNF). 
This is particularly promising since magnetism has been predicted 
in a variety of graphene nanostructures with zigzag (ZZ) edges\cite{sonNat444,rossierPRL99,ritterNM8,huJCP141,kabirPRB90,gangulyPRB95,valliPRB94} 
and experimentally found in gaphene nanoribbons\cite{taoNatPhys7,magdaNat514} 
and quantum dots\cite{sunQM2}.  
Even more interestingly the magnetic 
ordering surives up to room temperature.\cite{magdaNat514,valliPRB94} 

The combination of magnetism and QI leads to spin-polarized QI antiresonances, 
which can be used to obtain 
spin-filter\cite{yuSR6,ozakiPRB81,cocchiJCP133} 
and spin-valve\cite{chenNatComm4} effects, paving the way to the design 
of a novel generation of QI-tronic devices. 
In constrast with bulk graphene, where it was experimentally shown 
that a spin-resolved QI requires a magnetic field~\cite{lundebergNatPhys5}, 
the intrinsic magnetism of the ZGNF allows for spin filtering 
without any external field. 

We also demonstrate that the existence and the absence of QI features, 
and the effect of magnetic ordering, can be rationalized 
within a Green's function formalism in terms of basic symmetry properties. 
This observation underlines the generality of the phenomenon, which does not 
require a fine tuning of the geometry and of the model parameters and therefore
entails a huge potential  to predict and observe QI phenomena in a much wider class 
of nanoscale systems. Finally, the symmetry analysis also  leads us to propose a strategy 
for the electrostatic control of the spin-polarized transport.

In this work we theoreticaly study transport in ZGNFs 
within a tight-binding low-energy model describing the electrons 
delocalized in the $\pi$ orbital of graphene 
and interacting through a local (Hubbard) Coulomb repulsion $U$. 
We assume a perfect honeycomb lattice without distortions and a uniform 
hopping parameter $t$ between nearest-neighbor (NN) ions, 
which is representative of undoped nanoflakes with passivated edges.\cite{kabirPRB90} 
For the sake of formal simplicity we neglect the next-NN hopping  $t^\prime$, 
which is estimated\cite{kretininPRB88} as $t^\prime\!\approx\!0.1t$. 
This makes the model particle-hole symmetric, 
which helps our theoretical analysis. 
However our main results are robust to the inclusion 
of longer range hopping and, more generally, 
to the breaking of particle-hole symmetry, 
as we discuss explicitly in the Supporting Information (SI). 

We consider a device where the ZGNF, 
which acts as the correlated scattering region,
is 
contacted to two charge reservoirs (leads) at two of the ZZ edges. 
The spin-resolved transmission coefficient at equilibrium 
in the linear response regime reads \cite{landauerJRD1,meirPRL68,dattaETMS}
$T_{\sigma} = {\rm Tr} \Big( \Gamma^L G_{\sigma}^{a} 
                                        \Gamma^R G_{\sigma}^{r} 
                                 \Big)$,  where $G^{r(a)}_{\sigma}(\omega)$ is the interacting retarded (advanced) Green's function 
of the nanoflake in the presence of the leads. 
The matrix $\Gamma^{L/R}(\omega)$ is the spectral density 
of the left ($L$) and right ($R$) lead.

For our nanostructures we can recast the equation as 
\begin{equation}
\label{eq:tegij}
 T_{\sigma}(\omega) \!=\! \sum_{i}\sum_{j} 
                          \Gamma_{ii}^{L}(\omega) \Gamma_{jj}^{R}(\omega) 
                          |G^{r}_{ij \sigma}(\omega)|^2,  
\end{equation}  
where $i,j$ enumerate the edge sites connected to the $L,R$ leads, respectively,  
and select the corresponding matrix elements of $\Gamma^{L/R}(\omega)$. 
It is important to notice that Eq.~\eqref{eq:tegij} shows explicitly the relation 
between ballistic transport and the spectral properties of the scatterer, 
encoded in the real-space Green's function. 
This relation allows for a straightforward explanation of the QI patterns 
based on the symmetry properties of the spectrum. \cite{pedersenPRB90} 
It is indeed clear from Eq.~(\ref{eq:tegij}) that the simplest way to obtain a vanishing transmission
coefficient is that the Green's functions also vanish. This is indeed what we will find in the ZGNF 
when the leads are in a meta configuration. It is easy to realize that the link between zeros of the Green's
function and antiresonances is not limited to linear response
since any higher order term will be given by the product of a given number of Green's functions.

The explicit form of $\Gamma^{L/R}(\omega)$ depends on the details 
of the hybridization between the leads and the edges of the ZGNF. 
For simplicity, we assume a wide-band limit (WBL) approximation 
for the identical $L$ and $R$ leads, 
which results in a constant elements of the hybridization function, 
$\Gamma_{ii}^{L/R}(\omega)=\Gamma$ 
between the leads and the ZGNF. 
For the sake of definiteness we set $\Gamma/t=0.02$, 
but our assumptions are verified in a wide range of hybridization strengths, 
and our main results still remain valid when considering 
a dispersion for the leads beyond the WBL approximation (see SI). 

We evaluate the Green's function 
within the dynamical mean-field theory\cite{georgesRMP69} (DMFT) 
by means of a real-space extension\cite{snoekNJP10} 
which takes into account electronic correlations with a local, yet site-dependent,  approximation for the self-energy: 
$\Sigma_{ij\sigma}(\omega) = \Sigma_{ii\sigma}(\omega) \delta_{ij}$. 
This approach is suitable to describe the electronic 
and transport\cite{valliPRL104,valliPRB86,jacobPRB82,mazzaPRL117} properties 
of correlated systems lacking translational invariance 
in one or more spatial dimensions.\cite{schuelerEPJST226} 
Furthermore, DMFT can deal with broken-symmetry states\cite{georgesRMP69,snoekNJP10,amaricciPRA89,valliPRB94} 
and accurately captures the emergence of magnetic ordering in ZGNFs\cite{valliPRB94} 
which is crucial for the realization of the spin-filtering effect 
we shall discuss in this work. 
In the following we performs calculations at finite temperature $T/t=0.005$, 
but since the AF state of ZZ graphene nanostructures seems to be robust 
up to room temperature,\cite{valliPRB94,magdaNat514} 
upon varying $T$ we expect the main results presented here to remain valid.

\begin{figure}[t]
\includegraphics[width=0.5\textwidth, angle=0]{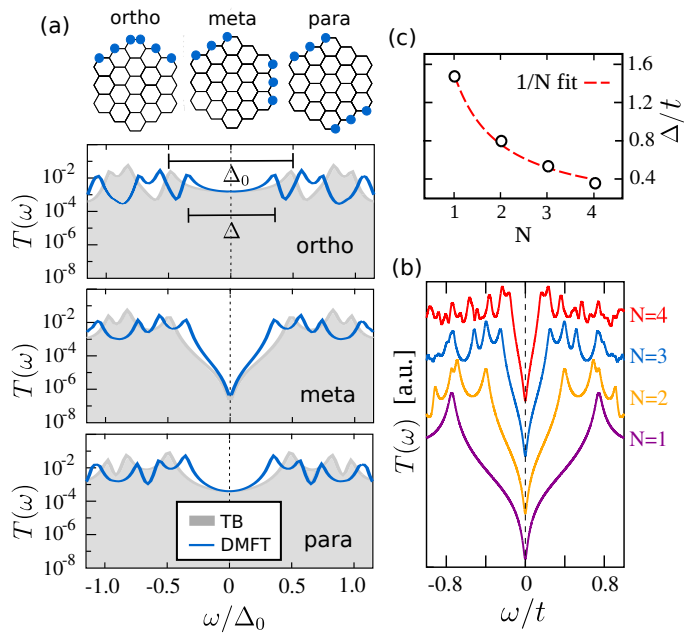}
\caption{{\b (a)} Transmission coefficient $T(\omega)$ 
in the ortho, meta, and para configurations for the $N\!=\!3$ ZGNF. 
In the upper panel each configuration is shown 
with cyan circles denoting the atoms connected to the leads. 
Electronic correlations within DMFT at $U/t\!=\!3.75$ 
reduce the spectral gap $\Delta<\Delta_0$ 
with respect to the tight-binding (TB) approximation. 
{\bf (b)} $T(\omega)$ in the meta configuration 
for ZGNF of different sizes $N\!=\!1\!-\!4$. 
The curves are shifted for improved readability. 
The QI antinode is stable with respect to system size, 
while {\bf (c)} the $\Delta$ is strongly size-dependent. }
\label{fig:Te_qi_graphene}
\end{figure}

In order to highlight and disentangle the role of magnetism and its interplay 
with the transport properties of ZGNFs, 
we start our discussion from paramagnetic (PM) solutions, 
where magnetism is explicitly inhibited, 
and we then compare with results in the magnetic phase, 
which becomes stable for moderate interaction strength $U/t$ 
(see the discussion in SI). 
We consider three different transport configurations 
in which the $L$ and $R$ leads are contacted to different ZZ edges. 
Depending on the relative position of the leads 
we refer to them as \emph{ortho},  \emph{meta}, and \emph{para} configurations 
(see Fig.~\ref{fig:Te_qi_graphene}), 
adopting the standard nomenclature used for hydrocarbon rings. 
In Fig.~\ref{fig:Te_qi_graphene}(a) we show 
the spin-independent zero-bias transmission coefficient $T(\omega)$ 
obtained within a PM calculation 
for a ZGNF with $N\!=\!3$ atoms at each ZZ edge. 
The most striking result is that  in the meta configuration $T(\omega)$ 
displays a transmission antiresonance at the Fermi level, 
which is a clear evidence of destructive QI, 
while the other configurations 
only show a reduction of transmission associated with the gap. 
This phenomenology is analogous to that of benzene 
molecular junctions\cite{solomonJCP129,cardamoneNL6,staffordNtech18,markussenNL10}. 
In Fig.~\ref{fig:Te_qi_graphene}(b) we show that QI effects  
are systematically observed in hexagonal ZGNFs of increasing size, 
ranging from $N\!=\!1$, corresponding to benzene, to $N\!=\!4$. 
The most relevant dependence on the increasing size is the reduction 
of the spectral gap,\cite{ritterNM8,huJCP141} 
shown in Fig.~\ref{fig:Te_qi_graphene}(c), 
which vanishes in bulk graphene, i.e. in the limit of $N\!\rightarrow\!\infty$.  
The spectral gap is also reduced by electronic correlations,\cite{valliPRB94} 
as demonstrated in Fig.~\ref{fig:Te_qi_graphene}(a), 
where we compare the tight-binding results for the non-interacting model 
with DMFT results 
for the $N\!=\!3$ ZGNF. 
Let us also notice, that upon reducing the gap, also the threshold value 
for the onset of the AF state is reduced, 
therefore favoring the occurrence of magnetism (see SI).  

In the case of a single benzene ring, destructive QI is indeed well understood 
in terms of the relative path difference of electrons 
propagating along the arms of the ring\cite{staffordNtech18}, 
an effect based essentially on symmetry which has been shown 
to survive the inclusion of electronic correlations.\cite{valliPRB86} 
However, the resilience of the QI patterns in the nanostructures we considered 
could not be easily anticipated. 
Due to the proliferation of transmission channels 
with the system size, QI effects in graphene nanostructures elude 
theoretical predictions obtained with different methods, i.e., 
molecular orbitals analysis\cite{solomonJCP129,solomonNC2,zhaoJPC146}, 
graphical rules\cite{stadlerNT15,markussenNL10,markussenPCCP13,markussenPRB89} 
or diagrammatic approaches,\cite{pedersenJPCC119} 
usually applied to linear\cite{solomonJACS130} 
or cyclic\cite{solomonJCP129,keNL8} $\pi$-conjugated molecules. 

It has been shown that QI effects can instead be related 
to the symmetries of the spectrum 
and the Coulson-Rushbrooke pairing theorem.\cite{nitaPSSRRL8,pedersenPRB90,zhaoJPC146} 
In this work we show that the analysis of the symmetries of the Hamiltonian 
allows to understand (and predict) destructive QI 
in very general terms, as a consequence of a vanishing Green's function. 
This way, we are also able to explain the interplay between QI and magnetism in ZGNFs, 
as we shall discuss in the following.

\begin{figure}[t]
\includegraphics[width=0.50\textwidth, angle=0]{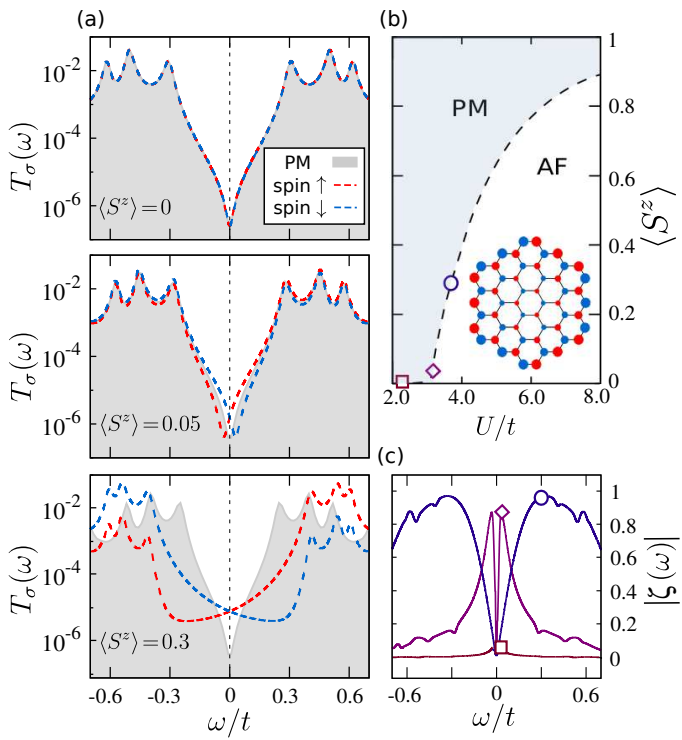}
\caption{{\bf (a)} Splitting of the spin-resolved transmission coefficient $T_{\sigma}(\omega)$ (red and cyan dashed lines) 
with respect to the unpolarized transmission in the PM state 
(light gray shaded area) for different values of the 
average staggered magnetization $\langle S^z \rangle$. 
{\bf (b)} $\langle S^z \rangle$ as a function of $U/t$ 
showing the onset of the fully-compensated AF state, 
with the real-space pattern represented in the inset. 
{\bf (c)} The maxima of the polarization $\zeta(\omega)$ are located 
in correspondence to the QI antinode at $\omega^{\textrm QI}$, 
which is proportional to $\langle S^z \rangle$  
(see the corresponding symbols in the two panels). }
\label{fig:Te_vsSz_multi}
\end{figure}

We now turn to the magnetic solutions 
which are expected to be stable for moderate interactions 
and have been identified in many theoretical\cite{sonNat444,rossierPRL99,ritterNM8,huJCP141,kabirPRB90,gangulyPRB95,valliPRB94}  and experimental\cite{taoNatPhys7,magdaNat514,sunQM2} studies
of graphene nanostructure with ZZ edges. 
While the details of the critical interaction strength 
to reach the magnetic state can depend on 
the specific nanostructure and its theoretical description 
(see the SI for some details), the 
very existence of magnetism appears as a solid and general evidence. 
For the typical $\pi$-conjugated systems in which QI effects 
have been recently addressed, magnetism is hardly relevant. 
Instead, intermediate-size graphene nanostructures with ZZ edges 
present the unique characteristic to host simultaneously 
QI effects, typical of molecular semiconductors, 
and a magnetic ordering, which is usually associated with bulk systems.
Consistent theoretical predictions \cite{sonNat444,rossierPRL99,ritterNM8,huJCP141,kabirPRB90,gangulyPRB95,valliPRB94} 
and experimental evidence\cite{taoNatPhys7,magdaNat514,sunQM2} 
suggest that graphene nanostructures with ZZ edges are prone 
to magnetic ordering, which survives up to room temperature.\cite{magdaNat514,valliPRB94} 
In particular, 
hexagonal ZGNFs display a fully-compensated (i.e., with zero net magnetic moment) 
antiferromagnetic (AF) order.\cite{rossierPRL99,kabirPRB90,gangulyPRB95,valliPRB94} 
The presence of the edges in the ZGNF determines an inhomogeneous 
spatial distribution of the ordered local magnetic moments 
$\langle S^z_i \rangle \!=\! \langle n_{i\uparrow}-n_{i\downarrow}\rangle$, 
which are significantly larger at the edges than in the bulk 
as a consequence of the reduced number of hopping channels 
for the edge sites.\cite{rossierPRL99,kabirPRB90,valliPRB94,feldnerPRB81} 
In hexagonal ZGNFs, all the atoms of a given edge of the hexagon 
belong to the same graphene sublattice, 
while neighboring edges are connected by an armchair (AC) defect, 
so that the local magnetic moments are aligned ferromagnetically 
within the same edge, and antiferromagnetically between neighboring edges, 
as shown in the inset of Fig.\ref{fig:Te_vsSz_multi}(b). 

In the ortho/para configurations the $L$ and $R$ leads 
are connected to atoms of different sublattices 
which host magnetic moments with opposite orientation. 
In these configurations the spin-degeneracy of $T_{\sigma}(\omega)$ 
is protected by symmetry, and the transport is not spin-polarized. 
Also from this point of view, the meta configuration results much more interesting, 
as the leads are connected to atoms of the same sublattice 
and with parallel magnetic moments. 
In Figs.~\ref{fig:Te_vsSz_multi}(a) we show 
the spin-resolved transmission coefficient $T_{\sigma}(\omega)$ 
in the meta configuration for different values 
of the average staggered magnetization of the nanoflake $\langle S^z \rangle$. 
In the presence of AF ordering the QI antiresonance is split 
symmetrically with respect to the Fermi level, 
resulting in spin-resolved QI features separated in energy and located 
at $\omega_\sigma^\textrm{QI} \ne 0$. 
The splitting is proportional to the average staggered magnetization, 
shown in \ref{fig:Te_vsSz_multi}(b) as a function of $U/t$. 
This is one of the main results of our work. 

To quantify this effect it is useful to consider  
the spin polarization of the transmission coefficient, defined as
\begin{equation} 
\label{eq:zeta}
\zeta(\omega) = \frac{T_{\uparrow}-T_{\downarrow}}{T_{\uparrow}+T_{\downarrow}}. 
\end{equation} 
As reported in Fig.~\ref{fig:Te_vsSz_multi}(c), 
the polarization at the Fermi level $\zeta(\omega\!=\!0)$ vanishes 
due to the symmetry around zero between the two 
spin-resolved transmission coefficients $T_{\sigma}(\omega)$. 
At finite frequency, the spin-polarization reaches instead values 
as high as $\zeta \approx 0.97$, i.e., nearly full polarization. 
Comparing Figs.~\ref{fig:Te_vsSz_multi}(a)~and~\ref{fig:Te_vsSz_multi}(c), 
we observe that the maxima of $|\zeta(\omega)|$ are located 
in correspondence of $\omega_{\sigma}^{\textrm QI}$, 
where the transmission in one of the spin channel 
is strongly suppressed by destructive QI. 
In turn, $\omega^{\textrm QI}_{\sigma}$ is proportional 
to $\langle S^z \rangle$, as indicated by the open symbols in
Figs.~\ref{fig:Te_vsSz_multi}(b) and \ref{fig:Te_vsSz_multi}(c). 
While the polarization at the Fermi level is zero due to symmetry, 
at finite external bias voltage 
it is possible to access a regime in which charge transport 
is dominated by one spin channel,  
achieving a nearly complete QI-assisted \emph{spin}-filtering effect. 
At the same time, it is also possible to select 
the \emph{spin}-polarization of the \emph{charge} current 
either by reversing the sign of the bias between the source and the drain, 
or connecting the leads in the meta configuration 
to edges of the other sublattice. 

The existence of destructive QI features in the transmission 
and their interplay with magnetism can be explained 
by analyzing the symmetries of the ZGNF within the Green's function formalism.
This represents a powerful tool to understand and predict destructive QI 
in complex nanostructures. 
At the same time, it also shows that the QI-assisted spin-filtering effect 
described above is a robust and generic feature of graphene nanostructures, 
and it does not depend on the details 
of the theoretical description of the system or of the leads. 

The Hamiltonian model describing ZGNFs has particle-hole symmetry 
(for $t^\prime\!=\!0$) and a chiral symmetry 
associated with the chemical equivalence between the two sublattices.
The two symmetries imply that the Green's function transforms in the following way 
\begin{equation}\label{eq:ph1}
 G^{*}_{ij\sigma}(\omega) = -\tau_i \tau_j G_{ij\sigma}(-\omega), 
\end{equation}
where $\tau_{i}\!=\!\pm 1$ denotes the chiral pseudospin at site $i$. 
In particular, the sign of the transformation in Eq.~(\ref{eq:ph1}) 
depends on whether $i$ and $j$ belong to the same ($\tau_i\tau_j\!=\!1$) 
or different ($\tau_i\tau_j\!=\!-1$) sublattices. 

This implies that any Green's function connecting two sites 
in the same sublattice will be a purely imaginary function for $\omega\!=\!0$, while
Green's functions connecting different sublattices 
will be purely real for $\omega\!=\!0$. 
These symmetries, combined with Eq. (\ref{eq:tegij}) 
imply the following relations for the transmission coefficients at the Fermi level
\begin{equation}\label{eq:T-PHortho}
T^{\rm ortho/para}_{\sigma}(0) = \Gamma^2 \sum_{i}\sum_{j} 
                                   \Big(\Re G_{ij\sigma}(0)\Big)^2,
\end{equation}
and 
\begin{equation}\label{eq:T-PHmeta}
T^{\rm meta}_{\sigma}(0)         = \Gamma^2 \sum_{i}\sum_{j} 
                                   \Big(\Im G_{ij\sigma}(0)\Big)^2.
\end{equation}
In the presence of a gap at the Fermi level, 
the imaginary part of the Green's function, 
which is proportional to the density of states, 
vanishes at $T\!=\!0$ and it is exponentially small at finite temperatures. 
This is the reason of the QI antinode in the meta configuration. 
On the other hand, the real part of the Green's function does not necessarily vanish, 
which is the reason why the ortho and para configurations 
display a depletion in the transmission function at the Fermi level 
but no destructive interference. 
This observation also explains the weak dependence of the QI 
on the size of the ZGNF, 
as increasing the systems size does not affect the symmetry of the nanostructure, 
but it just reduces the quantum confinement gap. 
Analogously, electronic correlations do not spoil the destructive QI 
as long as i) they do not close the energy gap, and ii) transport is phase-coherent. 
Finally, from  Eqs.~(\ref{eq:T-PHortho}-\ref{eq:T-PHmeta}) 
we deduce that the QI antiresonance is \emph{pinned} at the Fermi level 
in the presence of particle-hole and chiral symmetry.

In the AF state the $SU(2)$ spin invariance is broken, yet the Hamiltonian retains invariance under particle-hole transformation 
combined with spin inversion $\sigma \rightarrow \bar\sigma\!=\!-\sigma$. 
Then, for all configurations the following relation holds: 
\begin{equation}\label{eq:TSF}
 T_{\sigma}(\omega) = T_{\bar\sigma}(-\omega).
\end{equation}
As the magnetic sublattices coincide with the chiral sublattices, 
the spin inversion is equivalent to the inversion of the chiral
pseudospin, i.e., the exchange of A and B sublattice indices.  
Thus, for the meta configuration we obtain: 
\begin{equation}
 T^{\rm meta}_{\sigma AA}(\omega) = T^{\rm meta}_{\sigma BB}(-\omega),
\end{equation}
which combined with Eq.~(\ref{eq:TSF}) demonstrates the symmetry of the spin-resolved
transmission coefficients and the QI-assisted spin-filtering effect 
shown in Fig.~\ref{fig:Te_vsSz_multi}(a). 
In the ortho/para configurations instead, the spin-dependent transport 
is \emph{forbidden} by the symmetries of the model. 
Indeed, for these two cases the invariance under the particle-hole symmetry implies 
$T_{\sigma}(\omega)\!=\!T_{\sigma}(-\omega)$, 
that along with Eq.~(\ref{eq:TSF}) yields 
the spin-degeneracy of the transmission function: 
$T_{\sigma}(\omega)\!=\!T_{\bar \sigma}(\omega)$.

The above analysis clearly shows that the present results 
are much more general and fundamental
than their observation in the ZGNFs we addressed here. 
From Eq.~(\ref{eq:T-PHmeta}) it follows that QI antiresonance 
is biunivoquely associated to the existence 
of a zero of the Green's functions. 
Therefore similar results are expected for different molecules 
obeying similar symmetry properties which imply zeroes of the Green's function.

On the other hand, one might conclude that the existence of destructive QI 
is limited to molecules displaying such symmetries. 
In the following we show that this is not the case, 
i.e., that QI features persist when the particle-hole (see SI) 
or the chiral symmetries are broken.  

\begin{figure}
\includegraphics[width=0.5\textwidth, angle=0]{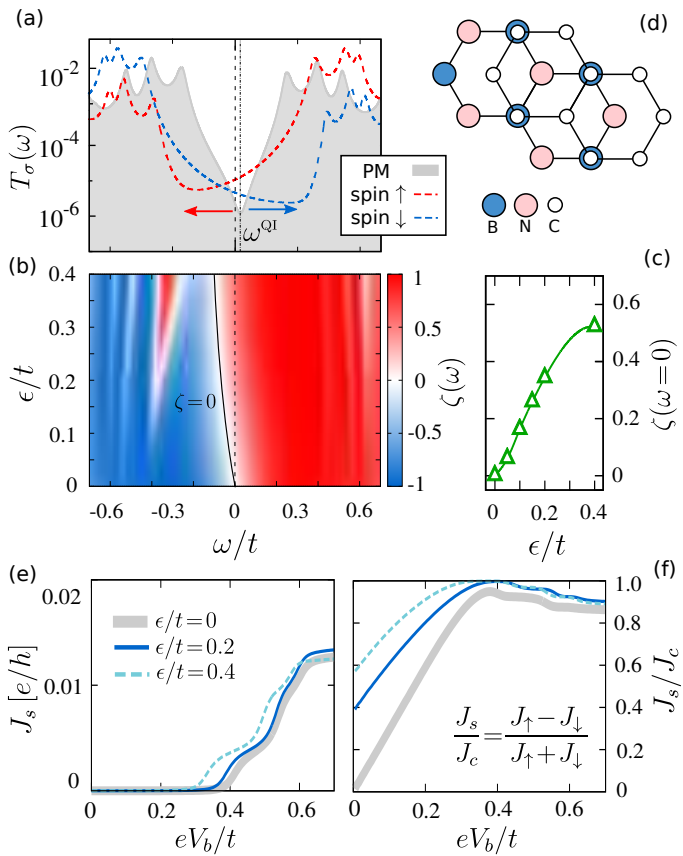}
\caption{{\bf (a)} Splitting of the spin-resolved 
transmission coefficient $T_{\sigma}(\omega)$ 
with respect to $\omega^{\textrm QI}\!\neq\!0$ 
for the graphene/h-BN heterostructure for $U/t\!=\!3.75$ and $\epsilon/t\!=\!0.2$. 
{\bf (b)} Heatmap of the spin-polarization 
of the transmission coefficient $\zeta(\omega,\epsilon)$.  
The zero of the polarization shifts away from the Fermi level (solid black line) 
due to the interaction with the substrate. 
{\bf (c)} A cut of $\zeta(\omega,\epsilon)$ at $\omega\!=\!0$ 
demonstrating the zero-bias spin-filtering effect. 
{\bf (d)} Schematic representation of the graphene/h-BN stacking 
in the stable configuration. 
{\bf (e-f)} Spin current $J_s$ and ratio $J_s/J_c$ between 
the spin and the charge current  
as a function of an external bias voltage $eV_b/t$ 
for different values of $\epsilon/t$, 
demonstrating the nearly-perfect spin-filtering effect. } 
\label{fig:ghBN}
\end{figure}

In particular we demonstrate how to achieve an electrostatic control 
of the spin polarization of the charge current 
by breaking the chiral symmetry of graphene lattice, 
e.g., by deposition on a suitable substrate. 
An optimal candidate for the realization of graphene devices 
is hexagonal boron nitride (h-BN).\cite{deanNatNanotech5} 
It has been shown within density-functional theory 
that for graphene/h-BN(0001) heterostructure 
in the most stable stacking configuration 
the two inequivalent C atoms of graphene are located respectively 
on top of the B atom and in the hollow position of h-BN.\cite{giovannettiPRB76} 
The corresponding equilibrium graphene/h-BN(0001) stacking 
is schematically represented in Fig.~\ref{fig:ghBN}(d). 
Due to the asymmetric absorption of C on the substrate  
the two graphene sublattices experience a different chemical environment, 
thus breaking the chiral symmetry and 
inducing a two-sublattice spatial charge modulation. 
The chiral symmetry-breaking field can be parameterized in terms of 
a staggered potential $\epsilon_i \!=\! \epsilon \tau_i$.\cite{skomskiMH563} 
Realistic estimates of the chiral symmetry-breaking field 
may vary from $\epsilon\!\approx\!0.1t\!-\!0.8t$, 
depending on the approximation.\cite{skomskiMH563,slawinskaPRB81}

Our calculations show that, if $\epsilon$ 
is smaller than the magnetic gap, 
the distribution of local magnetic moment $\langle S_i^z \rangle$ 
shown in Fig.~\ref{fig:Te_vsSz_multi}(b) 
is not affected by the charge modulation 
and that the AF pattern is preserved, 
although the moments are partially quenced (see SI). 
We find that the breaking of the chiral symmetry shifts the position of the 
QI antiresonance to a finite frequency $\omega_{\sigma}^{QI}$. 
Correspondingly, in the AF state $T_{\uparrow}(\omega)$ and $T_{\downarrow}(\omega)$ 
cross away from the Fermi level. 
This effect is demonstrated in Fig.~\ref{fig:ghBN}(a) for $\epsilon\!=\!0.2t$. 
We also observe that $T_{\uparrow}(\omega)$ and $T_{\downarrow}(\omega)$ 
are strongly asymmetric around the crossing, 
as a consequence of the charge imbalance between the two graphene sublattices, 
and the spin-polarization efficiency is not symmetric 
with respect to the sign of the bias voltage. 
The results in Fig.~\ref{fig:ghBN}(b-c) illustrate  
the effect of changing $\epsilon/t$ on the spin polarization $\zeta(\omega)$. 
The zero of $\zeta(\omega)$ continuously shifts toward negative frequencies 
by increasing the strength of $\epsilon/t$. 
This corresponds to a finite value of $\zeta(\omega\!=\!0)$ 
for any $\epsilon/t\!\neq\!0$. 
Remarkably, there is a wide energy window above the Fermi level 
near $\omega_{\sigma}^{\textrm QI}$ in which $\zeta(\omega)\!\approx\!1$ 
yielding a nearly complete spin-filtering effect.   

Direct consequences of the polarization of the transmission are
readily accessed by looking at the spin-resolved current 
\begin{equation}
 J_{\sigma} = \frac{e}{h} \int T_{\sigma}(\omega) \big[f_L(\omega)-f_R(\omega)\big] d\omega,
\end{equation}
where $f_{L/R}(\omega)$ is the Fermi distribution of the $L/R$ lead
at a chemical potential $\mu_{L}\!=\!\mu$ (equilibrium chemical potential) 
and $\mu_{R}\!=\!\mu-eV_b$, respectively, 
where $V_b$ is an external bias voltage.    
Due to the destructive QI in one of the spin channels, 
the driven current is highly spin polarized. 
As shown in Figs.~\ref{fig:ghBN}(e-f), 
by connecting the leads to ZZ edges with, e.g., a majority of spin down, 
one obtains $J_{\uparrow}\!\gg\!J_{\downarrow}$. The ratio between 
the spin current $J_s \!=\! J_{\uparrow}-J_{\downarrow}$ 
and the charge current $J_c \!=\! J_{\uparrow}+J_{\downarrow}$ 
increases with $eV_b/t$ and remains at $J_s/J_c>0.8$ 
for a wide range of bias values which becomes wider upon increasing $\epsilon/t$. 

It is worth to notice that when the chiral symmetry is broken  
the spin degeneracy is lifted also in the transmission coefficient 
of the ortho/para configurations. 
However, due to the lack of destructive QI, the resulting spin current
and the ratio $J_s/J_c$ are generally lower 
than in the meta configuration (see SI). 

The effect of the h-BN substrate on the ZGNF, encoded in the parameter $\epsilon$,  
can also be tuned electrostatically, 
e.g., by an external electric field which couples 
to the charge density modulation induced by the substrate.\cite{guanAPA123} 
This provides a systematic way to control the position of the QI antiresonances 
and the spin-filter efficiency. 
Experimentally, it was demonstrated that in bilayer graphene, 
where the inversion symmetry is broken by the AB (Bernal) stacking, 
the gap can be controlled by an external electric field.\cite{zhangNature459} 
In the experiment, it was shown that an electric field of $E=0.3$~V/\AA \
is able to induce a gap $\Delta\!\sim\!0.25$~eV ($\Delta\!\approx\!0.1t$). 
Such an electric field result in an electrostatic potential drop 
$\Delta V_{\textrm ext}\!=\!eEd$ along the stacking direction. 
With $e$ being the electric charge, 
$d\!=\!3.22$~\AA \ the equilibrium stacking distance obtained from 
density functional theory for graphene/h-BN bilayer,\cite{giovannettiPRB76} 
and taking $t\!\approx\!2.7$~eV for graphene,\cite{castronetoRMP81} 
it would correspond to a correction $\Delta V_{\textrm ext}\!\approx\!0.4t$ 
to the chiral symmetry-breaking field, 
$\epsilon_i\!=\!\epsilon\tau_i\!+\!\Delta V_{\textrm ext}$.

Finally, let us stress that the robustness of the destructive QI features  
poses a strong basis for the experimental realization 
of a graphene QI-assisted spin-filtering device. 
This is established on symmetry grounds, 
regardless on the theoretical details, e.g., 
of the description of the system and of the coupling to the leads. 
For instance, from Eq.~\eqref{eq:tegij} it is evident 
that the transmission antiresonance in the meta configuration 
is obtained from the contributions of independent channels. 
Based on the symmetry arguments (and in agreement with the numerical calculations) 
each channel individually display destructive QI. 
As a consequence, QI is naturally expected for any alternated hydrocarbon 
connected to leads through sites from the same sublattice. 
This is in agreement with recent observations in terms of 
the Coulson-Rushbrooke pairing theorem.\cite{zhaoJPC146,pedersenPRB90} 
In the case of graphene with ZZ edges, 
within each edge all the electrons available for binding 
to anchoring groups (or directly to the leads) 
belong to C atoms of the same sublattice. 
This suggests that the experimental observation of QI should be 
possible also in the case of imperfect contact 
between the leads and the ZZ edges. 
On the other hand, AC defects in a ZZ edge will open channels  
involving sites from different sublattices, 
whose contribution to the transmission can wash away the antiresonance. 
In this respect, the detection of QI antiresonances could also be exploited as an experimental probe for the purity of the ZZ edges. 

In summary, we have shown that hexagonal ZGNFs display 
clear signatures of quantum interference effects 
which strongly remind those of $\pi$-conjugated molecular junctions. 
In contrast with benzene rings, intermediate-size nanoflakes 
are also antiferromagnetically ordered, which couples
the QI effects with the spin polarization. 
In the meta configuration, 
it results in spin-dependent QI antiresonances separated in energy. 
This effect can be exploited for the realization of a spin-filter operating 
in the absence of an external magnetic field. 
We have shown in completely general terms the connection between 
the symmetries of the system and the existence of QI patterns, 
which also explains their interplay with the magnetization. 
Finally, we have demonstrated that the spin-polarization of the transmission 
can be controlled by an external parameter 
which breaks the chiral symmetry between the two sublattices of graphene. 
We propose deposition on a h-BN substrate as possible realization of this idea. 
Such a device would provide us with a controllable way to switch on and off 
the spin-filtering effect with a high degree of tunability, 
and should be within the reach of current technologies. 
The above analysis definitely calls for experimental evidence 
in support of our theoretical predictions. \\

We thank I.~H\'{a}gymasi R.~Stadler, and L.~Venkataraman for valuable discussions. 
We acknowledges financial support from MIUR PRIN 2015  (Prot. 2015C5SEJJ001) and SISSA/CNR project "Superconductivity, Ferroelectricity and Magnetism in bad metals" (Prot. 232/2015). AV acknowledges financial support from the Austrian Science Fund (FWF) 
through the Erwin Schr\"odinger fellowship J3890-N36. \\

\paragraph*{Supporting Information.} 
Emergence of edge magnetism within static- and dynamical mean-field theory 
as a function of system size; 
transport beyond the weak-hybridization regime 
and beyond the wide-band limit approximation; 
effect of particle-hole and chiral symmetry breaking (PDF).

\bibliographystyle{apsrev}
%


\pagebreak
\onecolumngrid

\setcounter{equation}{0}
\setcounter{figure}{0}
\setcounter{table}{0}
\setcounter{page}{1}
\makeatletter
\renewcommand{\thefigure}{S\arabic{figure}}
\renewcommand{\thepage}{S-\arabic{page}}

\newcommand*\mycommand[1]{\texttt{\emph{#1}}}

\begin{center}
  \textbf{\large Supporting Information: \\Quantum interference assisted spin filtering in graphene nanoflakes}\\[.2cm]
  A.~Valli,$^{1,*}$ A.~Amaricci,$^{1}$ V.~Brosco,$^1$ and M.~Capone$^1$\\[.1cm]
  {\itshape ${}^1$Scuola Internazionale Superiore di Studi Avanzati (SISSA) \\
CNR-IOM DEMOCRITOS, Istituto Officina dei Materiali, Consiglio Nazionale delle Ricerche, \\
Via Bonomea 265, 34136 Trieste, Italy}\\
\end{center}

\subsection*{Emergence of magnetism: gap engineering and quantum fluctuations} 

In this section we discuss the values of the model parameters 
used for the calculations reported in the manuscript and their impact on our results. 
In particular we focus on the conditions for the onset of magnetism in our nanoflakes.

The estimate of realistic interaction parameters for Hubbard-like modeling of solids 
is still a highly debated and very controversial issue in solid-state physics. 
The value of the bare Hubbard interaction among $\pi$-electrons was computed 
by Parr et al.\cite{parrJCP18} as $U\!=\!16.93$~eV. 
Assuming a value of $t\!\approx\!2.8$~eV for graphene,\cite{castronetoRMP81} 
it yields an estimate $U/t\!\approx\!6$. 
In order to obtain the effective $U$ for a Hubbard 
model, one has to consider the effect of screening. 
In a relatively small nanostructure we can expect a poorer screening 
with respect to a graphene sheet because of the finite-size 
gap due to the confined geometry. Therefore we assume a value of $U/t\!=\!3.75$.

We stress however that the message of the present manuscript 
does not rely on precise estimates of parameters. 
The basic  requirement to exploit quantum interference as a spin filter 
is that magnetism sets in for the values of the interaction we consider. 
The relevance of magnetism in graphene nanostructures with ZZ edges 
has been indeed predicted in a variety of papers and, most importantly, 
has been found experimentally in small nanoribbons, 
where the magnetic ordering survives up to room temperature, 
strongly supporting the relevance of our results.

\begin{figure}[b]
\includegraphics[width=0.9\textwidth, angle=0]{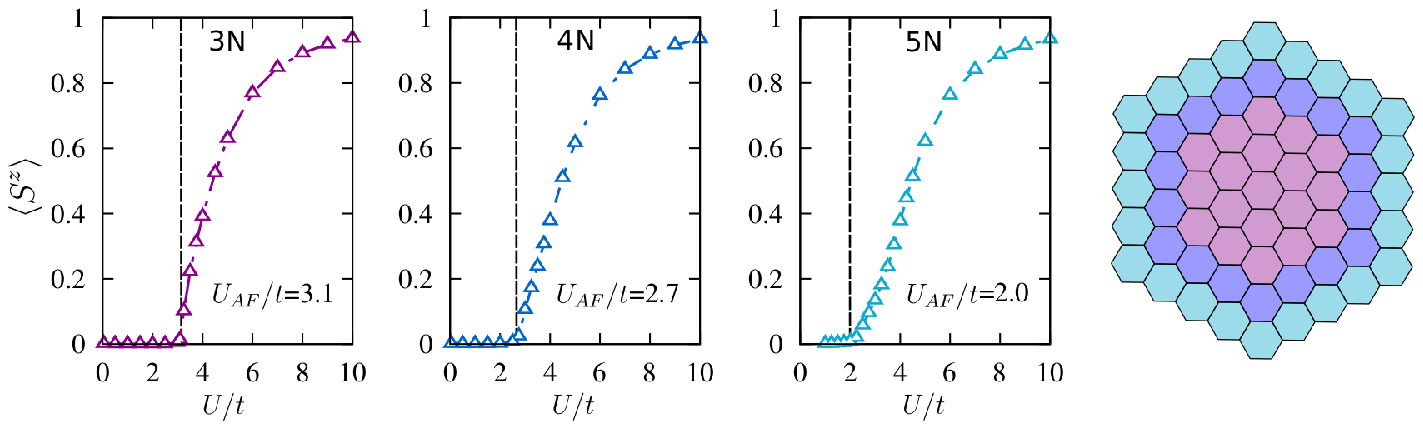}
\caption{Average magnetization $\langle S^z \rangle$ as a function of $U/t$ 
and corresponding numerical estimate of $U_{AF}/t$ (dahsed line) 
for nanoflakes with $N\!=\!3$, $N\!=\!4$, and $N\!=\!5$ edge atoms (right panel). }
\label{fig:mresU_size_fitUc}
\end{figure}

In a system with a finite spectral gap $\Delta$ the transition 
from the paramagnetic (PM) to the antiferromagnetic (AF) state inevitably 
sets in at a finite value of the Coulomb interaction ($U_{AF}$) which is mainly controlled by the amplitude of the gap.
In particular, in graphene nanoflakes the quantum confinement gap $\Delta$ 
decreases as the inverse of the linear size of the nanoflake $L$, 
as explicitly shown in the numerical calculations 
reported in Fig. 1(c) of the manuscript, 
in agreement with theoretical and experimental results 
in the literature.\cite{ritterNM8,huJCP141} 
As a result, also the value of $U_{AF}$ is suppressed 
upon increasing the size of the nanoflake.  
This is shown by our DMFT results  in Fig.~\ref{fig:mresU_size_fitUc}, 
where we show the average magnetization $\langle S^z \rangle$ of the nanoflake 
as a function of $U/t$, for nanoflakes of increasing size. 
Note that the value of $U_{AF}$ decreases 
from $U_{AF}/t\!\approx\!3.1$ to $U_{AF}/t\!\approx\!2.0$, 
upon increasing the linear size of the nanoflake from 
$L\!\approx\!14$~\AA~($3N$ nanoflake) to $L\!\approx\!25$~\AA~($5N$ nanoflake). 

It is important to stress, when comparing our results with previous literature, 
that our dynamical mean-field theory (DMFT) include quantum effects 
which tend to reduce the mean-field order parameter.
This leads to a substantial increases of  $U_{AF}$ with respect to static mean-field approaches, 
which completely neglect quantum fluctuation leading to small values of  $U_{AF}$ 
for graphene nanoflakes.\cite{rossierPRL99,gunlyckeNL7} 

To be concrete, we refer to calculations for a $3N$ hexagonal nanoflake,\cite{valliPRB94} 
where it is estimated $U_{AF}/t\!\approx\!2.0$ for static mean-field, 
and $U_{AF}/t\!\approx\!3.1$ for DMFT. 
Therefore, for a $5N$ hexagonal nanoflake, 
the static mean-field value of $U_{AF}/t\!\approx\!1.4$ 
indicated by Fernandez-Rossier~and~Palacio\cite{rossierPRL99} 
is perfectly compatible with our $U_{AF}/t\!\approx\!2.0$ DMFT estimate 
from Fig.~\ref{fig:mresU_size_fitUc}.

Independently on the specific value of $U/t$ chosen or the approximation employed, 
here we show that it is reasonably possible to realize magnetic nanoflakes 
also exploiting size engineering, 
i.e., lowering the value of $U_{AF}$ by reducing the quantum confinement gap.

\subsection*{Effect of the hybridization to the leads} 
\subsubsection*{Transport properties in the weak- and strong-hybridization regime} 
In the following we show the effect of the hybridization 
between the $3N$ nanoflake and the leads 
on the QI antinode in the meta configuration. 
In particular, we show the evolution of $T(\omega)$ from the weak- to the strong-hybridization regime. \\
The interacting Green's function of the isolated nanoflake 
(indicated with an empty circle, $G_{\circ}$) is defined as 
\begin{equation}
 G_{\circ}(\omega)\!=\!\big(\omega\!+\!\mu\!-\!\Sigma(\omega)\big)^{-1},
\end{equation}
where $\mu$ is the equilibrium chemical potential of the nanoflake 
and $\Sigma(\omega)$ is the dynamical self-energy matrix 
that takes into account electron-electron correlations. 
In the presence of the leads, the Green's function of the device 
(indicated with a filled circle, $G_{\bullet}$) 
is calculated by solving the Dyson following equation
\begin{equation}
G^{-1}_{\bullet}(\omega) = G^{-1}_{\circ}(\omega)
                             - \Sigma_L(\omega) - \Sigma_R(\omega), 
\end{equation}
where the leads are described by the embedding self-energy, defined as 
$\Gamma_{\alpha}(\omega)\!=\!-\frac{\imath}{2}
 \big(\Sigma^{r}_{\alpha}(\omega)\!-\!{\Sigma}^{a}_{\alpha}(\omega)\big)$, 
with $\alpha=L,R$ corresponding to the left ($L$) and right ($R$) leads. 

\begin{figure}[h]
\includegraphics[width=0.8\textwidth, angle=0]{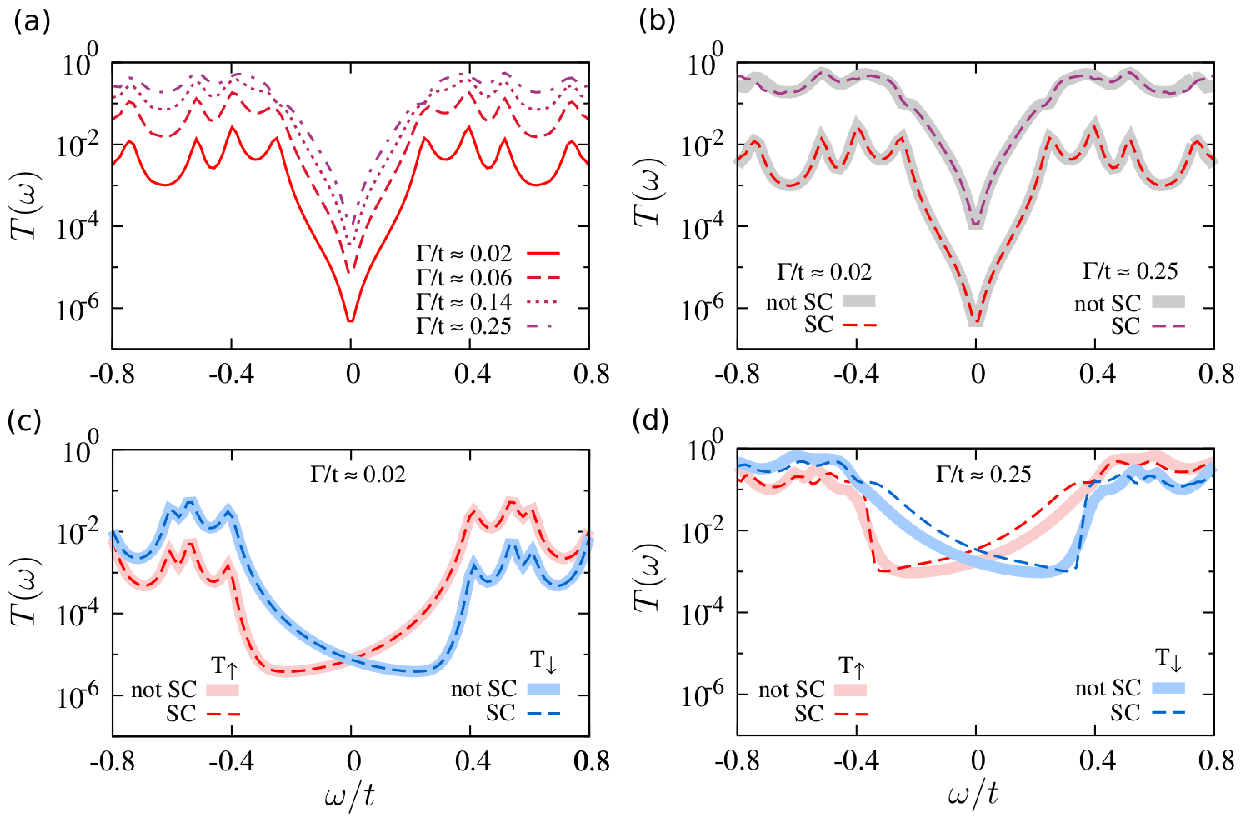}
\caption{$T(\omega)$ in the meta configuration for the $3N$ nanoflake. 
{\bf (a)} In the PM state: the QI antinode is robust 
by increasing the hybridization strength to the leads $\Gamma/t$. 
{\bf (b)} Comparison between non-self-consistent (shaded area) 
and fully self-consistent (dashed line) treatment of the leads 
on the transmission in the weak- ($\Gamma/t\!=\!0.02$) 
and strong-hybridization ($\Gamma/t\!=\!0.25$) regimes. 
{\bf (c-d)} Spin-resolved QI antinode in the AF state 
at $\Gamma/t\!=\!0.02$ and $\Gamma/t\!=\!0.25$, 
showing that the QI spin-filtering effect is robust against the values of $\Gamma/t$. }
\label{fig:effect_scleads}
\end{figure}

In Fig.~\ref{fig:effect_scleads}(a) we show 
the transmission coefficient $T(\omega)$ in the meta configuration 
(i.e., the one which exihibits a QI antinode) in the PM state. 
We show the evolution of $T(\omega)$ from the weak- ($\Gamma/t\!\approx\!0.02$) 
to the strong-hybridization ($\Gamma/t\!\approx\!0.25$) regimes. 
Upon increasing the hybridization we observe two main effects: 
i) the spectral features of the transmission become visibly broadened, and 
ii) the overall transmission increases (as the conductance $g\!\propto\!\Gamma^2$). 
However, due to destructive QI, the transmission at $\omega\!=\!0$ 
is strongly suppressed at any value of $\Gamma/t$, 
demonstrating the robustness of the QI features in all hybridization regimes.  

For the hexagonal nanoflake considered here $G_{\circ}(\omega)$ 
is invariant under the $C_{3}$ spatial rotation symmetry on the graphene plane. 
However, the presence of the leads breaks the $C_{3}$ rotational invariance 
and lowers the symmetry of the system (in all contact configurations). 
Hence, in order to evaluate the transmission, 
it is necessary to calculate $G_{\bullet}$ and all observables fully self-consistently.  
For the sake of completeness, in Fig.~\ref{fig:effect_scleads}(b) 
we also compare the meta transmission function $T(\omega)$ obtained 
in the weak- and strong-hybridization regimes, 
when the leads are taken into account self-consistently 
or when the Green's function of the isolated nanoflake 
is used to evaluate the transmission, 
showing that there is barely any difference in the resulting tranmission. 

In Figs.~\ref{fig:effect_scleads}(c,d) we show the same analysis in the AF state. 
Here, the leads perturb the distribution of the magnetic moments 
with respect to the isolated nanoflake, 
and the local magnetic moment $\langle S^z_i \rangle$ 
in the proximity of the corresponding edges 
are partially quenced (up to $30\%$ in the strong-hybridization limit). 
However, the splitting between the spin-resolved QI antinodes 
$\Delta\omega=\omega^{\textrm QI}_{\uparrow}-\omega^{\textrm QI}_{\downarrow}$ 
does not depend on the magnetic pattern within the nanoflake, 
but is controlled by the average staggered magnetization $\langle S^z \rangle$, 
which is, instead, only weakly affected by the leads. 
As a consequence, minor differences can be observed 
in the strong-hybridization limit 
(e.g., the position of $\omega^{\textrm QI}_{\sigma}$ change slighly) 
but the QI-assisted spin-filtering effect remains robust. 
Remarkably, while this effect is reasonable for large nanoflakes 
with lower surface-to-bulk ratio, 
we find this to be true even for the $3N$ nanoflake, 
which has a linear size of $L\!\approx\!14$~\AA, 
and a surface-to-bulk ratio of $1/3$. 
As a result, the spin-resolved QI antinode in the AF state 
is a robust feature of graphene nanoflake junctions, almost independently on size.

\subsubsection*{Transport properties beyond the wideband limit (WBL)} 
In the wide-nad limit (WBL), the embedding self-energy of the leads  
is a purely imaginary constant $\Sigma_{L/R}(\omega)\!=\!-\imath\Gamma$, 
which it contributes to the broadening of the many-body states of the nanoflake. 
A realistic hybridization to the leads would also include a real part 
which instead shifts the poles of the Green's function. 

\begin{figure}[h]
\includegraphics[width=0.8\textwidth, angle=0]{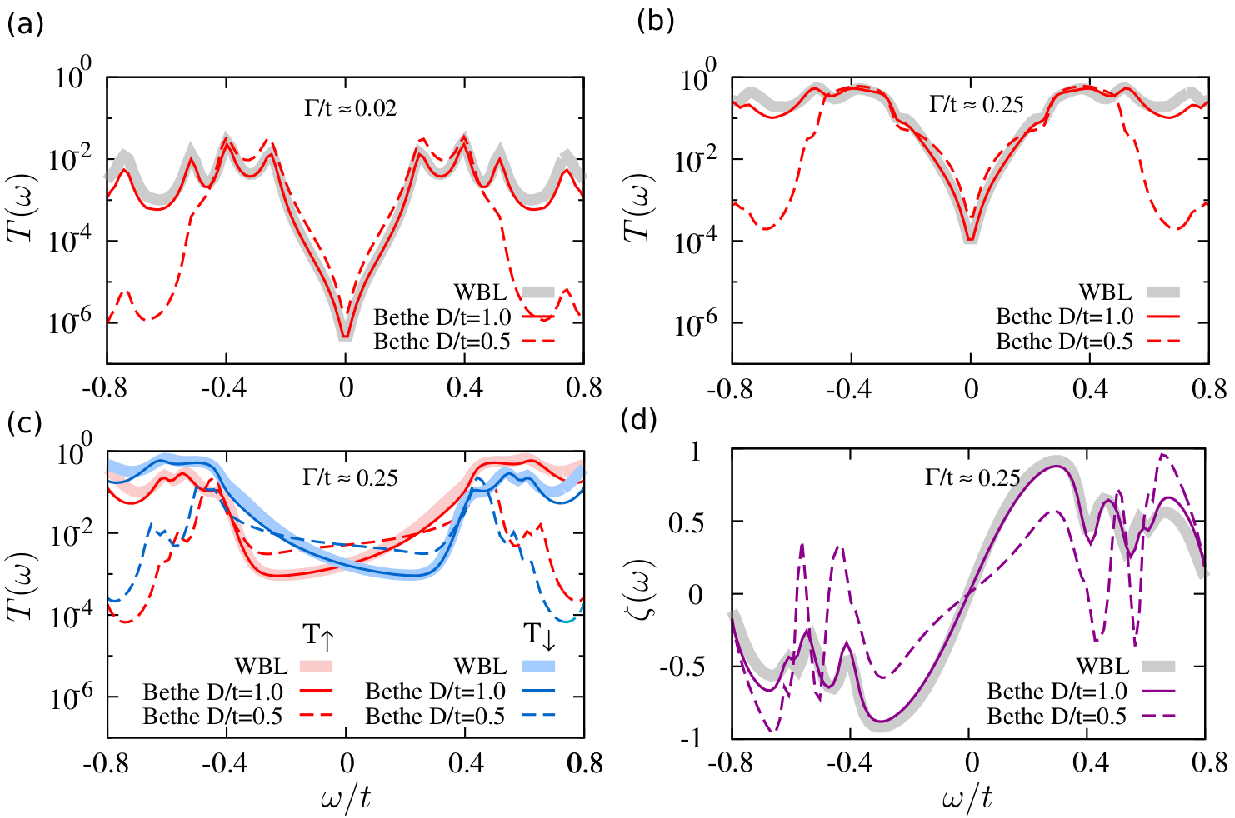}
\caption{$T(\omega)$ in the meta configuration for the $3N$ nanoflake 
comparing the wide-band limit (WBL) and a Bethe density of states for the leads.  
{\bf (a,b)} In the PM state: the QI antinode is robust 
by increasing the hybridization strength to the leads $\Gamma/t$.  
{\bf (c)} In the AF state: the QI antinote is robust. 
Some quantitative deviations can be observed 
with respect to the WBL in the extreme case 
in which the hybridization is strong ($\Gamma/t=0.25$) 
and the edge of the leads' DOS lies at the edge of the AF gap (i.e, for $D/t=0.5$). 
{\bf (d)} Spin-polarization of the transmission: 
even in the most unfavorable case the polarization is suppressed up to about $30\%$, 
but the spin-filtering effect survives. }
\label{fig:bethe}
\end{figure}

The applicability of the WBL in transport calculations 
has been discussed in details by Verzijl~et~al.,\cite{verzijlJCP138} 
concluding that WBL qualitatively reproduces 
the main features of the transmission and the bias-voltage dependence 
in cases where the transmission is dominated by the properties of the molecule. 
This sheds a positive light on our work, 
as the fundamental feature for the realization of the spin-filtering effect 
is indeed a property of the molecule, i.e., the destructive QI, 
and not the hybridization with the contacts. 

Nevertheless, in order to understand the possible effect of 
a realistic hybridization function on the transport properties, 
in the following we consider and explicitly take into account 
the effects of the leads beyond the WBL, 
and we present transport calculations 
with a semicircular (Bethe) density of states with a finite bandwidth $D$ for the leads. 
The corresponding embedding self-energy reads 
\begin{equation}
 \Sigma(\omega) = V^2 \frac{4}{\pi} \sum_{k} \frac{1}{\omega + \mu -\epsilon_k}
                      \sqrt{1-\Big(\frac{\epsilon_k}{D}\Big)^2}.
\end{equation}
In order to compare the Bethe lead with the WBL, 
we chose the hybridization parameter $V$ so that 
$\Gamma\!\equiv\!-V^2 \int_{-\infty}^{\infty}d\omega\Im\Sigma(\omega)/\pi$ 
for the Bethe leads is equal to the value of $\Gamma$ in the WBL. 
The results are shown in Figs.~\ref{fig:bethe}(a,b) for the PM state, 
and Figs.~\ref{fig:bethe}(c,d) for AF state. 
At low-energy, the calculations with the Bethe DOS reproduces 
the $T(\omega)$ of the WBL both in the 
weak- and in the strong-hybridization regimes 
(Figs.~\ref{fig:bethe}(a) and ~\ref{fig:bethe}(b), respectively). 
Obviously, the transmission is suppressed for $|\omega|>D$, 
but in the limit $D\!\gg\!t$, also the high-energy features would be recovered. 
The situation is more complex in the AF state, 
because the QI antinote is found at a finite frequency 
$|\omega^{\textrm QI}_{\sigma}|\!\propto\!\langle S^z \rangle$. 
In particular, if the edge of the Bethe DOS, 
where $\Re\Sigma_{L/R}(\omega)$ is the largest,  
is located in proximity of $\omega^{\textrm QI}_{\sigma}$, 
one can observe some deviations from the WBL, 
as in Fig.~\ref{fig:bethe}(c) for $D/t=0.5$. 
However, even in this case, which is the most unfavorable for the realization 
of the QI assisted spin-filtering effect, the spin polarization 
$\zeta(\omega)=(T_{\uparrow}-T_{\downarrow})/(T_{\uparrow}+T_{\downarrow})$ 
is found to be suppressed at most about $30\%$. 
This makes the efficiency of the device suboptimal but it does not destroy its spin-filtering properties. \\

\noindent We can conclude that the details of the DOS of the leads 
is generally irrelavant to the realization of the QI assisted spin-filtering effect. 
Together with the symmetry analysis that we discussed in the manuscript, 
these numerical calculations confirm that the phenomenon 
is a robust features of graphene junctions, 
and sheds promising lights on its experimental realization.

\subsection*{Effects of symmetry breaking on the destructive QI}
In the manuscript we show that the QI features of graphene nanostructures 
can be understood in terms of the symmetries of the Hamiltonian, 
which establish their robustness and generality. 
However, the existence of QI features does \emph{not} rely on those symmetries, 
and that indeed the QI antiresonances appear in the transmission 
even when the symmetries are lifted.  
Here we show that the QI properties survive even if those symmetries are broken. 
We focus on two case: i) when the particle-hole symmetry is broken 
by the presence of hoppings beyond nearest neighbors (NN), and 
ii) when the chiral symmetry of graphene is broken by the presence of a substrate.  

\subsubsection*{Structural electron-hole asymmetry: $t^{\prime}\!\neq\!0$}
As a matter of fact, in actual graphene nanostructures 
the particle-hole symmetry of NN hopping tight-binding Hamiltonian, 
as considered above, 
is expected to be broken by hopping processes beyond NN 
or by lattice deformations. 
Here we consider the case in which we include 
next-NN hopping $t^\prime\!\neq\!0$ in the tight-binding Hamiltonian. 

\begin{figure*}[t]
\includegraphics[width=1.0\textwidth, angle=0]{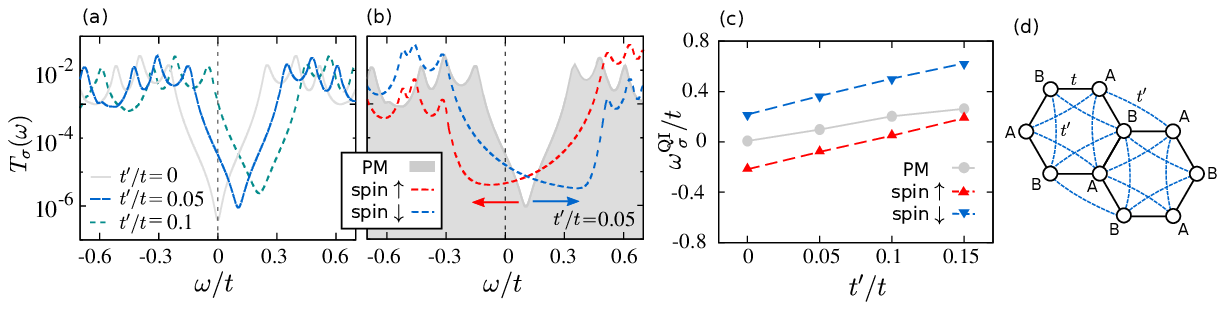}
\caption{{\bf (a-b)} Transmisson coefficient in the PM and AF state 
for different values of the particle-hole symmetry breaking $t^\prime/t$ 
for $\Gamma/t=0.02$ and $U/t\!=\!3.75$. 
{\bf (c)} Shift of the QI antinode $\omega^{\textrm QI}$ as a function $t^\prime/t$. 
{\bf (d)} While NN hopping $t$ connect atoms from different sublattices (AB), 
the next-NN hopping $t^\prime$ connects atoms within the same sublattice (AA or BB). } 
\label{fig:Te_effect_ttp}
\end{figure*}

In Fig.~\ref{fig:Te_effect_ttp}(a) we show explicitly that $t^{\prime}$ 
does not destroy the QI antiresonance, 
but it shifts the frequency $\omega^{\rm QI}$ 
(at which the destructive interference takes place) 
at finite energy. 
We can understand this result observing that $t^\prime$ 
shifts the DOS, so that the chemical potential no longer lies in the middle of the gap, 
but slightly below (for positive $t^\prime/t$). 
However, if the spectrum at $t^{\prime}\!=\!0$ was symmetric around $\omega\!=\!0$, 
it still possesses a \emph{near symmetry} 
around a shifted energy $\omega^{\textrm QI}\!\neq\!0$. 
In particular, here $\omega^{\textrm QI}\!\propto\!t^{\prime}/t$. 
This, in turn, results in a zero of the Green's function at the same energy, 
away from the Fermi level. 

It is also important to notice that a relatively small next-NN hopping, 
besides driving the system away from half-filling, 
does not destroy the ordered AF state due to the presence of the quantum confinement gap. 
In the AF state we observe that 
$T_{\uparrow}(\omega)$ and $T_{\downarrow}(\omega)$  
are split around the antiresonance energy $\omega^{\textrm QI}$, 
as shown in Fig.~\ref{fig:Te_effect_ttp}(b) for $t^\prime/t\!=\!0.05$. 

While a finite $t^\prime$ is a simple conceptual handle 
to tune the position of the QI antiresonance, 
this parameter is not easily tuned in actual materials. 
Similar effects can be obtained using deformations of the lattice structure 
induced by applying, e.g., strain. 
The possibility to mechanically control QI has been recently demonstrated 
experimentally for $\pi$-stacked dimers exhibiting destructive QI.\cite{frisendaNatChem8} 
Yet, in complex nanostructures it may be difficult to achieve a 
precise control over the local lattice deformations 
and ultimately on the position of the QI antiresonance.

\subsubsection*{Chiral symmetry breaking: graphene/h-BN(0001)}
Upon deposition of graphene on hexagonal boron-nitride (h-BN) 
the two inequivalent $A$ and $B$ sublattices experience 
a different chemical environment due to asymmetric absorption on the substrate. 
This effect can be modeled by the following  substrate Hamiltonian\cite{skomskiMH563} 
\begin{equation}
\label{eq:Hcsb}
 H_{\rm sub}= -\sum_{i} \big( \epsilon_A n_{iA} +\epsilon_{B} n_{iB} \big), 
\end{equation}
where $n_{iA(B)}$ is the electron density operators at site $i$ on sublattice $A(B)$ 
and $\epsilon_A\!=\!-\epsilon_B\!=\!\epsilon$ is the parameter 
that measures the degree of chiral symmetry-breaking. 

The term in Eq.~\eqref{eq:Hcsb} induces a charge modulation between the two sublattices, as the occupation of all sites deviates from half-filling. 
This is quantified by the charge-density wave 
order parameter $\Delta_{\rm CDW}=N_C^{-1}\sum_{i}\big(n_{iA}-n_{iB}\big)$, 
with $N_C$ the number of C atoms in the nanostructure. 
As a consequence, also the local magnetic moments $\langle S^z_i \rangle$ 
are partially quenced, and the staggered magnetization 
$\langle S^z \rangle \!=\!
  N_C^{-1}\sum_{i}\big(\langle S^z_{iA}\rangle-\langle S^z_{iB} \rangle\big)$ 
is reduced. 
However, the AF pattern of the magnetization is nevertheless preserved. 

\begin{figure}
\includegraphics[width=0.32\textwidth, angle=0]{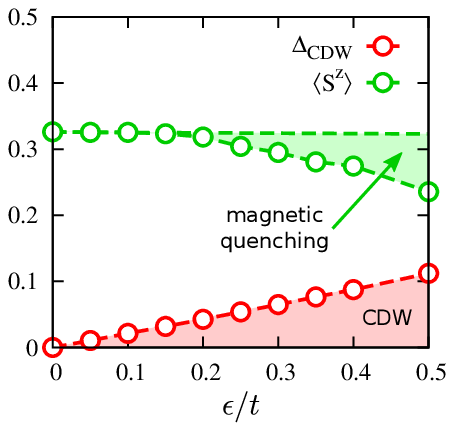}
\caption{Amplitde of the charge-density wave $\Delta_{\rm CDW}$ order parameter 
and the staggered magnetization $\langle S^z \rangle$ 
for the isolated $3N$ nanoflake at $U/t=3.75$. }
\label{fig:csb}
\end{figure}

In Fig.~\ref{fig:csb} we show the amplitude of $\Delta_{\rm CDW}$ 
and $\langle S^z \rangle$ for the isolated $3N$ nanoflake at $U/t=3.75$, 
as a function of the hybridization between the nanoflake and the substrate. 

\begin{figure}
\includegraphics[width=0.9\textwidth, angle=0]{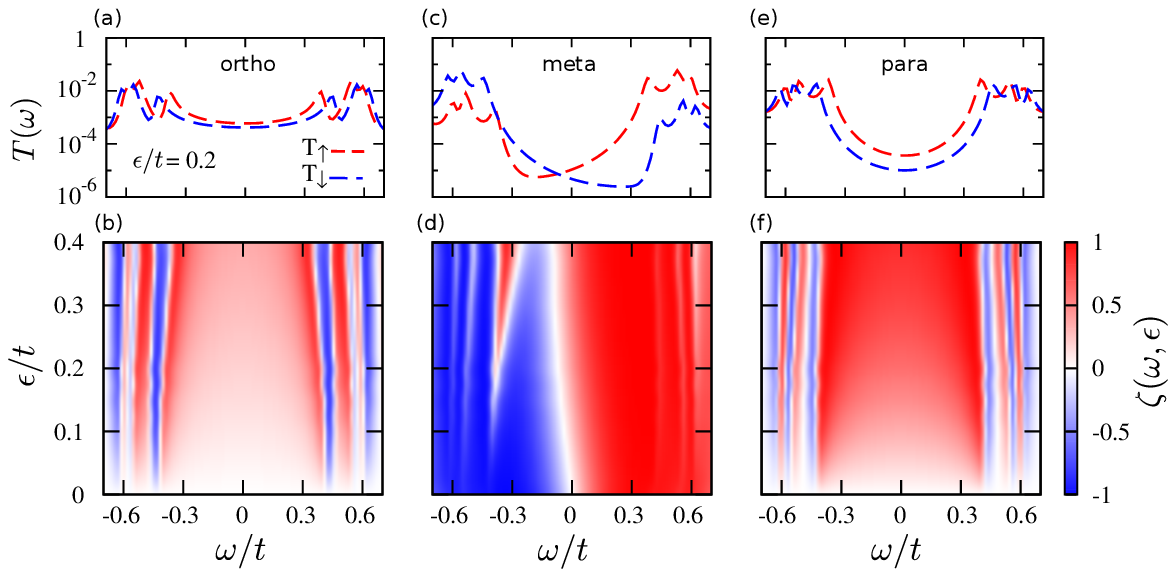}
\caption{Transmission coefficient for $\epsilon/t=0.2$ 
and heatmap of the spin-polarization $\zeta(\omega,\epsilon)$ 
in the ortho {\bf (a-b)}, meta {\bf(c-d)}, and para {\bf (e-f)} 
configurations for the $3N$ nanoflake, for $\Gamma/t=0.02$ and $U/t=3.75$. }
\label{fig:Te_map_omp}
\end{figure}

In order to compare the effects of the chiral symmetry-breaking field $\epsilon$ 
in all contact configurations, 
in Fig.~\ref{fig:Te_map_omp} we show the ortho, meta, and para 
transmission coefficient $T_{\sigma}(\omega)$ of the $3N$ nanoflake device. 
In general, for the ortho and para configurations 
(which do not display destructive QI) 
the polarization in the proximity of the Fermi level 
is lower than in the meta configuration. 
For each spin polarization $\sigma$, the transmission coefficient 
in the ortho and para configurations 
fulfills the condition $T_{\sigma}(\omega)=T_{\sigma}(-\omega)$ 
(while the substrate breaks the relation between spin $\uparrow$ and $\downarrow$). 
As a result, there is a finite $\Delta T(0)=T_{\uparrow}(0)-T_{\downarrow}(0)$ 
but the local minimum of the transmission is pinned at the Fermi level. 
Eventually, this is the reason why the corresponding spin-current $J_s$ 
is lower than in meta configuration. 

Therefore, we can conclude that the suppression of the transmission 
in one spin channel, due to spin-resolved destructive QI 
achieved in the meta configuration, 
is fundamental in order to obtain an efficient spin filter. 
In particular, QI-assisted spin-filtering is more effective 
than any polarization induced in the ortho or para configurations.

\bibliographystyle{apsrev}

\begin{thebibliography}{62}%
\makeatletter
\providecommand \@ifxundefined [1]{%
 \@ifx{#1\undefined}
}%
\providecommand \@ifnum [1]{%
 \ifnum #1\expandafter \@firstoftwo
 \else \expandafter \@secondoftwo
 \fi
}%
\providecommand \@ifx [1]{%
 \ifx #1\expandafter \@firstoftwo
 \else \expandafter \@secondoftwo
 \fi
}%
\providecommand \natexlab [1]{#1}%
\providecommand \enquote  [1]{``#1''}%
\providecommand \bibnamefont  [1]{#1}%
\providecommand \bibfnamefont [1]{#1}%
\providecommand \citenamefont [1]{#1}%
\providecommand \href@noop [0]{\@secondoftwo}%
\providecommand \href [0]{\begingroup \@sanitize@url \@href}%
\providecommand \@href[1]{\@@startlink{#1}\@@href}%
\providecommand \@@href[1]{\endgroup#1\@@endlink}%
\providecommand \@sanitize@url [0]{\catcode `\\12\catcode `\$12\catcode
  `\&12\catcode `\#12\catcode `\^12\catcode `\_12\catcode `\%12\relax}%
\providecommand \@@startlink[1]{}%
\providecommand \@@endlink[0]{}%
\providecommand \url  [0]{\begingroup\@sanitize@url \@url }%
\providecommand \@url [1]{\endgroup\@href {#1}{\urlprefix }}%
\providecommand \urlprefix  [0]{URL }%
\providecommand \Eprint [0]{\href }%
\providecommand \doibase [0]{http://dx.doi.org/}%
\providecommand \selectlanguage [0]{\@gobble}%
\providecommand \bibinfo  [0]{\@secondoftwo}%
\providecommand \bibfield  [0]{\@secondoftwo}%
\providecommand \translation [1]{[#1]}%
\providecommand \BibitemOpen [0]{}%
\providecommand \bibitemStop [0]{}%
\providecommand \bibitemNoStop [0]{.\EOS\space}%
\providecommand \EOS [0]{\spacefactor3000\relax}%
\providecommand \BibitemShut  [1]{\csname bibitem#1\endcsname}%
\let\auto@bib@innerbib\@empty
\bibitem [{\citenamefont {Cardamone}\ \emph {et~al.}(2006)\citenamefont
  {Cardamone}, \citenamefont {Stafford},\ and\ \citenamefont
  {Mazumdar}}]{cardamoneNL6}%
  \BibitemOpen
  \bibfield  {author} {\bibinfo {author} {\bibfnamefont {D.~M.}\ \bibnamefont
  {Cardamone}}, \bibinfo {author} {\bibfnamefont {C.~A.}\ \bibnamefont
  {Stafford}}, \ and\ \bibinfo {author} {\bibfnamefont {S.}~\bibnamefont
  {Mazumdar}},\ }\href@noop {} {\bibfield  {journal} {\bibinfo  {journal} {Nano
  Lett.}\ }\textbf {\bibinfo {volume} {6}},\ \bibinfo {pages} {2422} (\bibinfo
  {year} {2006})}\BibitemShut {NoStop}%
\bibitem [{\citenamefont {Stafford}\ \emph {et~al.}(2007)\citenamefont
  {Stafford}, \citenamefont {Cardamone},\ and\ \citenamefont
  {Mazumdar}}]{staffordNtech18}%
  \BibitemOpen
  \bibfield  {author} {\bibinfo {author} {\bibfnamefont {C.~A.}\ \bibnamefont
  {Stafford}}, \bibinfo {author} {\bibfnamefont {D.~M.}\ \bibnamefont
  {Cardamone}}, \ and\ \bibinfo {author} {\bibfnamefont {S.}~\bibnamefont
  {Mazumdar}},\ }\href@noop {} {\bibfield  {journal} {\bibinfo  {journal}
  {Nanotechnology}\ }\textbf {\bibinfo {volume} {18}},\ \bibinfo {pages}
  {424014} (\bibinfo {year} {2007})}\BibitemShut {NoStop}%
\bibitem [{\citenamefont {Solomon}\ \emph
  {et~al.}(2008{\natexlab{a}})\citenamefont {Solomon}, \citenamefont {Andrews},
  \citenamefont {Hansen}, \citenamefont {Goldsmith}, \citenamefont
  {Wasielewski}, \citenamefont {Duyne},\ and\ \citenamefont
  {Ratner}}]{solomonJCP129}%
  \BibitemOpen
  \bibfield  {author} {\bibinfo {author} {\bibfnamefont {G.~C.}\ \bibnamefont
  {Solomon}}, \bibinfo {author} {\bibfnamefont {D.~Q.}\ \bibnamefont
  {Andrews}}, \bibinfo {author} {\bibfnamefont {T.}~\bibnamefont {Hansen}},
  \bibinfo {author} {\bibfnamefont {R.~H.}\ \bibnamefont {Goldsmith}}, \bibinfo
  {author} {\bibfnamefont {M.~R.}\ \bibnamefont {Wasielewski}}, \bibinfo
  {author} {\bibfnamefont {R.~P.~V.}\ \bibnamefont {Duyne}}, \ and\ \bibinfo
  {author} {\bibfnamefont {M.~A.}\ \bibnamefont {Ratner}},\ }\href@noop {}
  {\bibfield  {journal} {\bibinfo  {journal} {J. Chem. Phys.}\ }\textbf
  {\bibinfo {volume} {129}},\ \bibinfo {pages} {054701} (\bibinfo {year}
  {2008}{\natexlab{a}})}\BibitemShut {NoStop}%
\bibitem [{\citenamefont {Stadler}\ \emph {et~al.}(2003)\citenamefont
  {Stadler}, \citenamefont {Forshaw},\ and\ \citenamefont
  {Joachim}}]{stadlerNT14}%
  \BibitemOpen
  \bibfield  {author} {\bibinfo {author} {\bibfnamefont {R.}~\bibnamefont
  {Stadler}}, \bibinfo {author} {\bibfnamefont {M.}~\bibnamefont {Forshaw}}, \
  and\ \bibinfo {author} {\bibfnamefont {C.}~\bibnamefont {Joachim}},\
  }\href@noop {} {\bibfield  {journal} {\bibinfo  {journal} {Nanotechnology}\
  }\textbf {\bibinfo {volume} {14}},\ \bibinfo {pages} {138} (\bibinfo {year}
  {2003})}\BibitemShut {NoStop}%
\bibitem [{\citenamefont {Stadler}\ \emph {et~al.}(2004)\citenamefont
  {Stadler}, \citenamefont {Ami}, \citenamefont {Forshaw},\ and\ \citenamefont
  {Joachim}}]{stadlerNT15}%
  \BibitemOpen
  \bibfield  {author} {\bibinfo {author} {\bibfnamefont {R.}~\bibnamefont
  {Stadler}}, \bibinfo {author} {\bibfnamefont {S.}~\bibnamefont {Ami}},
  \bibinfo {author} {\bibfnamefont {M.}~\bibnamefont {Forshaw}}, \ and\
  \bibinfo {author} {\bibfnamefont {C.}~\bibnamefont {Joachim}},\ }\href@noop
  {} {\bibfield  {journal} {\bibinfo  {journal} {Nanotechnology}\ }\textbf
  {\bibinfo {volume} {15}},\ \bibinfo {pages} {S115} (\bibinfo {year}
  {2004})}\BibitemShut {NoStop}%
\bibitem [{\citenamefont {Solomon}\ \emph {et~al.}(2010)\citenamefont
  {Solomon}, \citenamefont {Herrmann}, \citenamefont {Herrmann}, \citenamefont
  {Hansen}, \citenamefont {Mujica},\ and\ \citenamefont {Ratner}}]{solomonNC2}%
  \BibitemOpen
  \bibfield  {author} {\bibinfo {author} {\bibfnamefont {G.~C.}\ \bibnamefont
  {Solomon}}, \bibinfo {author} {\bibfnamefont {C.}~\bibnamefont {Herrmann}},
  \bibinfo {author} {\bibfnamefont {C.}~\bibnamefont {Herrmann}}, \bibinfo
  {author} {\bibfnamefont {T.}~\bibnamefont {Hansen}}, \bibinfo {author}
  {\bibfnamefont {V.}~\bibnamefont {Mujica}}, \ and\ \bibinfo {author}
  {\bibfnamefont {M.~A.}\ \bibnamefont {Ratner}},\ }\href@noop {} {\bibfield
  {journal} {\bibinfo  {journal} {Nat. Chem.}\ }\textbf {\bibinfo {volume}
  {2}},\ \bibinfo {pages} {223} (\bibinfo {year} {2010})}\BibitemShut {NoStop}%
\bibitem [{\citenamefont {Saha}\ \emph {et~al.}(2010)\citenamefont {Saha},
  \citenamefont {Nikoli\'{c}}, \citenamefont {Meunier}, \citenamefont {Lu}, ,\
  and\ \citenamefont {Bernholc}}]{sahaPRL105}%
  \BibitemOpen
  \bibfield  {author} {\bibinfo {author} {\bibfnamefont {K.~K.}\ \bibnamefont
  {Saha}}, \bibinfo {author} {\bibfnamefont {B.~K.}\ \bibnamefont
  {Nikoli\'{c}}}, \bibinfo {author} {\bibfnamefont {V.}~\bibnamefont
  {Meunier}}, \bibinfo {author} {\bibfnamefont {W.}~\bibnamefont {Lu}}, , \
  and\ \bibinfo {author} {\bibfnamefont {J.}~\bibnamefont {Bernholc}},\
  }\href@noop {} {\bibfield  {journal} {\bibinfo  {journal} {Phys. Rev. Lett.}\
  }\textbf {\bibinfo {volume} {105}},\ \bibinfo {pages} {236803} (\bibinfo
  {year} {2010})}\BibitemShut {NoStop}%
\bibitem [{\citenamefont {Markussen}\ \emph {et~al.}(2010)\citenamefont
  {Markussen}, \citenamefont {Stadler},\ and\ \citenamefont
  {Thygesen}}]{markussenNL10}%
  \BibitemOpen
  \bibfield  {author} {\bibinfo {author} {\bibfnamefont {T.}~\bibnamefont
  {Markussen}}, \bibinfo {author} {\bibfnamefont {R.}~\bibnamefont {Stadler}},
  \ and\ \bibinfo {author} {\bibfnamefont {K.~S.}\ \bibnamefont {Thygesen}},\
  }\href@noop {} {\bibfield  {journal} {\bibinfo  {journal} {Nano Lett.}\
  }\textbf {\bibinfo {volume} {10}},\ \bibinfo {pages} {4260} (\bibinfo {year}
  {2010})}\BibitemShut {NoStop}%
\bibitem [{\citenamefont {Markussen}\ \emph {et~al.}(2011)\citenamefont
  {Markussen}, \citenamefont {Stadler},\ and\ \citenamefont
  {Thygesen}}]{markussenPCCP13}%
  \BibitemOpen
  \bibfield  {author} {\bibinfo {author} {\bibfnamefont {T.}~\bibnamefont
  {Markussen}}, \bibinfo {author} {\bibfnamefont {R.}~\bibnamefont {Stadler}},
  \ and\ \bibinfo {author} {\bibfnamefont {K.~S.}\ \bibnamefont {Thygesen}},\
  }\href@noop {} {\bibfield  {journal} {\bibinfo  {journal} {Phys. Chem. Chem.
  Phys.}\ }\textbf {\bibinfo {volume} {13}},\ \bibinfo {pages} {14311}
  (\bibinfo {year} {2011})}\BibitemShut {NoStop}%
\bibitem [{\citenamefont {Markussen}\ and\ \citenamefont
  {Thygesen}(2014)}]{markussenPRB89}%
  \BibitemOpen
  \bibfield  {author} {\bibinfo {author} {\bibfnamefont {T.}~\bibnamefont
  {Markussen}}\ and\ \bibinfo {author} {\bibfnamefont {K.~S.}\ \bibnamefont
  {Thygesen}},\ }\href@noop {} {\bibfield  {journal} {\bibinfo  {journal}
  {Phys. Rev. B}\ }\textbf {\bibinfo {volume} {89}},\ \bibinfo {pages} {085420}
  (\bibinfo {year} {2014})}\BibitemShut {NoStop}%
\bibitem [{\citenamefont {Gerlich}\ \emph {et~al.}(2011)\citenamefont
  {Gerlich}, \citenamefont {Eibenberger}, \citenamefont {Tomandl},
  \citenamefont {Nimmrichter}, \citenamefont {Hornberger}, \citenamefont
  {Fagan}, \citenamefont {Tüxen}, \citenamefont {Mayor},\ and\ \citenamefont
  {Arndt}}]{gerlichNatComm2}%
  \BibitemOpen
  \bibfield  {author} {\bibinfo {author} {\bibfnamefont {S.}~\bibnamefont
  {Gerlich}}, \bibinfo {author} {\bibfnamefont {S.}~\bibnamefont
  {Eibenberger}}, \bibinfo {author} {\bibfnamefont {M.}~\bibnamefont
  {Tomandl}}, \bibinfo {author} {\bibfnamefont {S.}~\bibnamefont
  {Nimmrichter}}, \bibinfo {author} {\bibfnamefont {K.}~\bibnamefont
  {Hornberger}}, \bibinfo {author} {\bibfnamefont {P.~J.}\ \bibnamefont
  {Fagan}}, \bibinfo {author} {\bibfnamefont {J.}~\bibnamefont {Tüxen}},
  \bibinfo {author} {\bibfnamefont {M.}~\bibnamefont {Mayor}}, \ and\ \bibinfo
  {author} {\bibfnamefont {M.}~\bibnamefont {Arndt}},\ }\href@noop {}
  {\bibfield  {journal} {\bibinfo  {journal} {Nat. Comm.}\ }\textbf {\bibinfo
  {volume} {2}},\ \bibinfo {pages} {263} (\bibinfo {year} {2011})}\BibitemShut
  {NoStop}%
\bibitem [{\citenamefont {Richert}\ \emph {et~al.}(2017)\citenamefont
  {Richert}, \citenamefont {Cremers}, \citenamefont {Kuprov}, \citenamefont
  {Peeks}, \citenamefont {Anderson},\ and\ \citenamefont
  {Timmel}}]{richertNatComm8}%
  \BibitemOpen
  \bibfield  {author} {\bibinfo {author} {\bibfnamefont {S.}~\bibnamefont
  {Richert}}, \bibinfo {author} {\bibfnamefont {J.}~\bibnamefont {Cremers}},
  \bibinfo {author} {\bibfnamefont {I.}~\bibnamefont {Kuprov}}, \bibinfo
  {author} {\bibfnamefont {M.~D.}\ \bibnamefont {Peeks}}, \bibinfo {author}
  {\bibfnamefont {H.~L.}\ \bibnamefont {Anderson}}, \ and\ \bibinfo {author}
  {\bibfnamefont {C.~R.}\ \bibnamefont {Timmel}},\ }\href@noop {} {\bibfield
  {journal} {\bibinfo  {journal} {Nat. Commun.}\ }\textbf {\bibinfo {volume}
  {8}},\ \bibinfo {pages} {8:14842} (\bibinfo {year} {2017})}\BibitemShut
  {NoStop}%
\bibitem [{\citenamefont {Ke}\ \emph {et~al.}(2008)\citenamefont {Ke},
  \citenamefont {Yang},\ and\ \citenamefont {Baranger}}]{keNL8}%
  \BibitemOpen
  \bibfield  {author} {\bibinfo {author} {\bibfnamefont {S.-H.}\ \bibnamefont
  {Ke}}, \bibinfo {author} {\bibfnamefont {W.}~\bibnamefont {Yang}}, \ and\
  \bibinfo {author} {\bibfnamefont {H.~U.}\ \bibnamefont {Baranger}},\
  }\href@noop {} {\bibfield  {journal} {\bibinfo  {journal} {Nano Lett.}\
  }\textbf {\bibinfo {volume} {8}},\ \bibinfo {pages} {3257} (\bibinfo {year}
  {2008})}\BibitemShut {NoStop}%
\bibitem [{\citenamefont {Frisenda}\ \emph {et~al.}(2016)\citenamefont
  {Frisenda}, \citenamefont {Janssen}, \citenamefont {Grozema}, \citenamefont
  {der Zant},\ and\ \citenamefont {Renaud}}]{frisendaNatChem8}%
  \BibitemOpen
  \bibfield  {author} {\bibinfo {author} {\bibfnamefont {R.}~\bibnamefont
  {Frisenda}}, \bibinfo {author} {\bibfnamefont {V.~A.~E.~C.}\ \bibnamefont
  {Janssen}}, \bibinfo {author} {\bibfnamefont {F.~C.}\ \bibnamefont
  {Grozema}}, \bibinfo {author} {\bibfnamefont {H.~S.}\ \bibnamefont {der
  Zant}}, \ and\ \bibinfo {author} {\bibfnamefont {N.}~\bibnamefont {Renaud}},\
  }\href@noop {} {\bibfield  {journal} {\bibinfo  {journal} {Nat. Chem.}\
  }\textbf {\bibinfo {volume} {8}},\ \bibinfo {pages} {1099} (\bibinfo {year}
  {2016})}\BibitemShut {NoStop}%
\bibitem [{\citenamefont {Pedersen}\ \emph {et~al.}(2015)\citenamefont
  {Pedersen}, \citenamefont {Borges}, \citenamefont {Hedegard}, \citenamefont
  {Solomon},\ and\ \citenamefont {Strange}}]{pedersenJPCC119}%
  \BibitemOpen
  \bibfield  {author} {\bibinfo {author} {\bibfnamefont {K.~G.~L.}\
  \bibnamefont {Pedersen}}, \bibinfo {author} {\bibfnamefont {A.}~\bibnamefont
  {Borges}}, \bibinfo {author} {\bibfnamefont {P.}~\bibnamefont {Hedegard}},
  \bibinfo {author} {\bibfnamefont {G.~C.}\ \bibnamefont {Solomon}}, \ and\
  \bibinfo {author} {\bibfnamefont {M.}~\bibnamefont {Strange}},\ }\href@noop
  {} {\bibfield  {journal} {\bibinfo  {journal} {J. Phys. Chem. C.}\ }\textbf
  {\bibinfo {volume} {119}},\ \bibinfo {pages} {26919} (\bibinfo {year}
  {2015})}\BibitemShut {NoStop}%
\bibitem [{\citenamefont {Xia}\ \emph {et~al.}(2014)\citenamefont {Xia},
  \citenamefont {Capozzi}, \citenamefont {Wei}, \citenamefont {Strange},
  \citenamefont {Batra}, \citenamefont {Moreno}, \citenamefont {Amir},
  \citenamefont {Amir}, \citenamefont {Solomon}, \citenamefont {Venkataraman},\
  and\ \citenamefont {Campos}}]{xiaNL14}%
  \BibitemOpen
  \bibfield  {author} {\bibinfo {author} {\bibfnamefont {J.}~\bibnamefont
  {Xia}}, \bibinfo {author} {\bibfnamefont {B.}~\bibnamefont {Capozzi}},
  \bibinfo {author} {\bibfnamefont {S.}~\bibnamefont {Wei}}, \bibinfo {author}
  {\bibfnamefont {M.}~\bibnamefont {Strange}}, \bibinfo {author} {\bibfnamefont
  {A.}~\bibnamefont {Batra}}, \bibinfo {author} {\bibfnamefont {J.~R.}\
  \bibnamefont {Moreno}}, \bibinfo {author} {\bibfnamefont {R.~J.}\
  \bibnamefont {Amir}}, \bibinfo {author} {\bibfnamefont {E.}~\bibnamefont
  {Amir}}, \bibinfo {author} {\bibfnamefont {G.~C.}\ \bibnamefont {Solomon}},
  \bibinfo {author} {\bibfnamefont {L.}~\bibnamefont {Venkataraman}}, \ and\
  \bibinfo {author} {\bibfnamefont {L.~M.}\ \bibnamefont {Campos}},\
  }\href@noop {} {\bibfield  {journal} {\bibinfo  {journal} {Nano Lett.}\
  }\textbf {\bibinfo {volume} {14}},\ \bibinfo {pages} {2941} (\bibinfo {year}
  {2014})}\BibitemShut {NoStop}%
\bibitem [{\citenamefont {Stadler}(2015)}]{stadlerNL15}%
  \BibitemOpen
  \bibfield  {author} {\bibinfo {author} {\bibfnamefont {R.}~\bibnamefont
  {Stadler}},\ }\href@noop {} {\bibfield  {journal} {\bibinfo  {journal} {Nano
  Lett.}\ }\textbf {\bibinfo {volume} {15}},\ \bibinfo {pages} {7175} (\bibinfo
  {year} {2015})}\BibitemShut {NoStop}%
\bibitem [{\citenamefont {Strange}\ \emph {et~al.}(2015)\citenamefont
  {Strange}, \citenamefont {Solomon}, \citenamefont {Venkataraman},\ and\
  \citenamefont {Campos}}]{strangeNL15}%
  \BibitemOpen
  \bibfield  {author} {\bibinfo {author} {\bibfnamefont {M.}~\bibnamefont
  {Strange}}, \bibinfo {author} {\bibfnamefont {G.~C.}\ \bibnamefont
  {Solomon}}, \bibinfo {author} {\bibfnamefont {L.}~\bibnamefont
  {Venkataraman}}, \ and\ \bibinfo {author} {\bibfnamefont {L.~M.}\
  \bibnamefont {Campos}},\ }\href@noop {} {\bibfield  {journal} {\bibinfo
  {journal} {Nano Lett.}\ }\textbf {\bibinfo {volume} {15}},\ \bibinfo {pages}
  {7177} (\bibinfo {year} {2015})}\BibitemShut {NoStop}%
\bibitem [{\citenamefont {Zhao}\ \emph {et~al.}(2017)\citenamefont {Zhao},
  \citenamefont {Geskin},\ and\ \citenamefont {Stadler}}]{zhaoJPC146}%
  \BibitemOpen
  \bibfield  {author} {\bibinfo {author} {\bibfnamefont {X.}~\bibnamefont
  {Zhao}}, \bibinfo {author} {\bibfnamefont {V.}~\bibnamefont {Geskin}}, \ and\
  \bibinfo {author} {\bibfnamefont {R.}~\bibnamefont {Stadler}},\ }\href@noop
  {} {\bibfield  {journal} {\bibinfo  {journal} {J. Chem. Phys.}\ }\textbf
  {\bibinfo {volume} {146}},\ \bibinfo {pages} {092308} (\bibinfo {year}
  {2017})}\BibitemShut {NoStop}%
\bibitem [{\citenamefont {Pedersen}\ \emph {et~al.}(2014)\citenamefont
  {Pedersen}, \citenamefont {Strange}, \citenamefont {.Leijnse}, \citenamefont
  {Hedegard}, \citenamefont {Solomon},\ and\ \citenamefont
  {Pasaske}}]{pedersenPRB90}%
  \BibitemOpen
  \bibfield  {author} {\bibinfo {author} {\bibfnamefont {K.~G.~L.}\
  \bibnamefont {Pedersen}}, \bibinfo {author} {\bibfnamefont {M.}~\bibnamefont
  {Strange}}, \bibinfo {author} {\bibfnamefont {M.}~\bibnamefont {.Leijnse}},
  \bibinfo {author} {\bibfnamefont {P.}~\bibnamefont {Hedegard}}, \bibinfo
  {author} {\bibfnamefont {G.~C.}\ \bibnamefont {Solomon}}, \ and\ \bibinfo
  {author} {\bibfnamefont {J.}~\bibnamefont {Pasaske}},\ }\href@noop {}
  {\bibfield  {journal} {\bibinfo  {journal} {Phys. Rev. B}\ }\textbf {\bibinfo
  {volume} {90}},\ \bibinfo {pages} {25413} (\bibinfo {year}
  {2014})}\BibitemShut {NoStop}%
\bibitem [{\citenamefont {Mun\'{a}rriz}\ \emph {et~al.}(2011)\citenamefont
  {Mun\'{a}rriz}, \citenamefont {Dom\'{i}nguez-Adame},\ and\ \citenamefont
  {Malyshev}}]{munarrizNanotech22}%
  \BibitemOpen
  \bibfield  {author} {\bibinfo {author} {\bibfnamefont {J.}~\bibnamefont
  {Mun\'{a}rriz}}, \bibinfo {author} {\bibfnamefont {F.}~\bibnamefont
  {Dom\'{i}nguez-Adame}}, \ and\ \bibinfo {author} {\bibfnamefont {A.~V.}\
  \bibnamefont {Malyshev}},\ }\href@noop {} {\bibfield  {journal} {\bibinfo
  {journal} {Nanotechnology}\ }\textbf {\bibinfo {volume} {22}},\ \bibinfo
  {pages} {365201} (\bibinfo {year} {2011})}\BibitemShut {NoStop}%
\bibitem [{\citenamefont {Nozaki}\ \emph {et~al.}(2015)\citenamefont {Nozaki},
  \citenamefont {Lokamani}, \citenamefont {Santana-Bonilla}, \citenamefont
  {Dianat}, \citenamefont {Gutierrez}, ,\ and\ \citenamefont
  {Cuniberti}}]{nozakiJPCL6}%
  \BibitemOpen
  \bibfield  {author} {\bibinfo {author} {\bibfnamefont {D.}~\bibnamefont
  {Nozaki}}, \bibinfo {author} {\bibnamefont {Lokamani}}, \bibinfo {author}
  {\bibfnamefont {A.}~\bibnamefont {Santana-Bonilla}}, \bibinfo {author}
  {\bibfnamefont {A.}~\bibnamefont {Dianat}}, \bibinfo {author} {\bibfnamefont
  {R.}~\bibnamefont {Gutierrez}}, , \ and\ \bibinfo {author} {\bibfnamefont
  {G.}~\bibnamefont {Cuniberti}},\ }\href@noop {} {\bibfield  {journal}
  {\bibinfo  {journal} {J. Phys. Chem. Lett.}\ }\textbf {\bibinfo {volume}
  {6}},\ \bibinfo {pages} {3950} (\bibinfo {year} {2015})}\BibitemShut
  {NoStop}%
\bibitem [{\citenamefont {Ni\cb{t}\u{a}}\ \emph {et~al.}(2014)\citenamefont
  {Ni\cb{t}\u{a}}, \citenamefont {\cb{T}olea},\ and\ \citenamefont
  {Ostahie}}]{nitaPSSRRL8}%
  \BibitemOpen
  \bibfield  {author} {\bibinfo {author} {\bibfnamefont {M.}~\bibnamefont
  {Ni\cb{t}\u{a}}}, \bibinfo {author} {\bibfnamefont {M.}~\bibnamefont
  {\cb{T}olea}}, \ and\ \bibinfo {author} {\bibfnamefont {R.}~\bibnamefont
  {Ostahie}},\ }\href@noop {} {\bibfield  {journal} {\bibinfo  {journal} {Phys.
  Stauts Solidi RRL}\ }\textbf {\bibinfo {volume} {8}},\ \bibinfo {pages} {790}
  (\bibinfo {year} {2014})}\BibitemShut {NoStop}%
\bibitem [{\citenamefont {Geng}\ \emph {et~al.}(2015)\citenamefont {Geng},
  \citenamefont {Sangtarash}, \citenamefont {Huang}, \citenamefont {Sadeghi},
  \citenamefont {Fu}, \citenamefont {Hong}, \citenamefont {Wandlowski},
  \citenamefont {Decurtins}, \citenamefont {Lambert},\ and\ \citenamefont
  {Liu}}]{gengJACS137}%
  \BibitemOpen
  \bibfield  {author} {\bibinfo {author} {\bibfnamefont {Y.}~\bibnamefont
  {Geng}}, \bibinfo {author} {\bibfnamefont {S.}~\bibnamefont {Sangtarash}},
  \bibinfo {author} {\bibfnamefont {C.}~\bibnamefont {Huang}}, \bibinfo
  {author} {\bibfnamefont {H.}~\bibnamefont {Sadeghi}}, \bibinfo {author}
  {\bibfnamefont {Y.}~\bibnamefont {Fu}}, \bibinfo {author} {\bibfnamefont
  {W.}~\bibnamefont {Hong}}, \bibinfo {author} {\bibfnamefont {T.}~\bibnamefont
  {Wandlowski}}, \bibinfo {author} {\bibfnamefont {S.}~\bibnamefont
  {Decurtins}}, \bibinfo {author} {\bibfnamefont {C.~J.}\ \bibnamefont
  {Lambert}}, \ and\ \bibinfo {author} {\bibfnamefont {S.-X.}\ \bibnamefont
  {Liu}},\ }\href@noop {} {\bibfield  {journal} {\bibinfo  {journal} {J. Am.
  Chem. Soc.}\ }\textbf {\bibinfo {volume} {137}},\ \bibinfo {pages}
  {4469–4476} (\bibinfo {year} {2015})}\BibitemShut {NoStop}%
\bibitem [{\citenamefont {Sangtarash}\ \emph {et~al.}(2016)\citenamefont
  {Sangtarash}, \citenamefont {Sadeghi},\ and\ \citenamefont
  {Lambert}}]{sangtarashNanoscale8}%
  \BibitemOpen
  \bibfield  {author} {\bibinfo {author} {\bibfnamefont {S.}~\bibnamefont
  {Sangtarash}}, \bibinfo {author} {\bibfnamefont {H.}~\bibnamefont {Sadeghi}},
  \ and\ \bibinfo {author} {\bibfnamefont {C.~J.}\ \bibnamefont {Lambert}},\
  }\href@noop {} {\bibfield  {journal} {\bibinfo  {journal} {Nanoscale}\
  }\textbf {\bibinfo {volume} {8}},\ \bibinfo {pages} {13199} (\bibinfo {year}
  {2016})}\BibitemShut {NoStop}%
\bibitem [{\citenamefont {Gehring}\ \emph {et~al.}(2016)\citenamefont
  {Gehring}, \citenamefont {Sadeghi}, \citenamefont {Sangtarash}, \citenamefont
  {Lau}, \citenamefont {J.~Liu}, \citenamefont {Warner}, \citenamefont
  {Lambert}, \citenamefont {Briggs},\ and\ \citenamefont {Mol}}]{gehringNL16}%
  \BibitemOpen
  \bibfield  {author} {\bibinfo {author} {\bibfnamefont {P.}~\bibnamefont
  {Gehring}}, \bibinfo {author} {\bibfnamefont {H.}~\bibnamefont {Sadeghi}},
  \bibinfo {author} {\bibfnamefont {S.}~\bibnamefont {Sangtarash}}, \bibinfo
  {author} {\bibfnamefont {C.~S.}\ \bibnamefont {Lau}}, \bibinfo {author}
  {\bibfnamefont {A.~A.}\ \bibnamefont {J.~Liu}}, \bibinfo {author}
  {\bibfnamefont {J.~H.}\ \bibnamefont {Warner}}, \bibinfo {author}
  {\bibfnamefont {C.~J.}\ \bibnamefont {Lambert}}, \bibinfo {author}
  {\bibfnamefont {G.~A.~D.}\ \bibnamefont {Briggs}}, \ and\ \bibinfo {author}
  {\bibfnamefont {J.~A.}\ \bibnamefont {Mol}},\ }\href@noop {} {\bibfield
  {journal} {\bibinfo  {journal} {Nano Lett.}\ }\textbf {\bibinfo {volume}
  {16}},\ \bibinfo {pages} {4210} (\bibinfo {year} {2016})}\BibitemShut
  {NoStop}%
\bibitem [{\citenamefont {Son}\ \emph {et~al.}(2006)\citenamefont {Son},
  \citenamefont {Cohen},\ and\ \citenamefont {Louie}}]{sonNat444}%
  \BibitemOpen
  \bibfield  {author} {\bibinfo {author} {\bibfnamefont {Y.~W.}\ \bibnamefont
  {Son}}, \bibinfo {author} {\bibfnamefont {M.~L.}\ \bibnamefont {Cohen}}, \
  and\ \bibinfo {author} {\bibfnamefont {S.~G.}\ \bibnamefont {Louie}},\
  }\href@noop {} {\bibfield  {journal} {\bibinfo  {journal} {Nature}\ }\textbf
  {\bibinfo {volume} {444}},\ \bibinfo {pages} {347} (\bibinfo {year}
  {2006})}\BibitemShut {NoStop}%
\bibitem [{\citenamefont {Fern\'{a}ndez-Rossier}\ and\ \citenamefont
  {Palacios}(2007)}]{rossierPRL99}%
  \BibitemOpen
  \bibfield  {author} {\bibinfo {author} {\bibfnamefont {J.}~\bibnamefont
  {Fern\'{a}ndez-Rossier}}\ and\ \bibinfo {author} {\bibfnamefont {J.~J.}\
  \bibnamefont {Palacios}},\ }\href@noop {} {\bibfield  {journal} {\bibinfo
  {journal} {Phys. Rev. Lett.}\ }\textbf {\bibinfo {volume} {99}},\ \bibinfo
  {pages} {177204} (\bibinfo {year} {2007})}\BibitemShut {NoStop}%
\bibitem [{\citenamefont {Ritter}\ and\ \citenamefont
  {Lyding}(2009)}]{ritterNM8}%
  \BibitemOpen
  \bibfield  {author} {\bibinfo {author} {\bibfnamefont {K.~A.}\ \bibnamefont
  {Ritter}}\ and\ \bibinfo {author} {\bibfnamefont {J.~W.}\ \bibnamefont
  {Lyding}},\ }\href@noop {} {\bibfield  {journal} {\bibinfo  {journal} {Nat.
  Mat.}\ }\textbf {\bibinfo {volume} {8}},\ \bibinfo {pages} {235} (\bibinfo
  {year} {2009})}\BibitemShut {NoStop}%
\bibitem [{\citenamefont {Hu}\ \emph {et~al.}(2014)\citenamefont {Hu},
  \citenamefont {Lin}, \citenamefont {Yang},\ and\ \citenamefont
  {Yang}}]{huJCP141}%
  \BibitemOpen
  \bibfield  {author} {\bibinfo {author} {\bibfnamefont {W.}~\bibnamefont
  {Hu}}, \bibinfo {author} {\bibfnamefont {L.}~\bibnamefont {Lin}}, \bibinfo
  {author} {\bibfnamefont {C.}~\bibnamefont {Yang}}, \ and\ \bibinfo {author}
  {\bibfnamefont {J.}~\bibnamefont {Yang}},\ }\href@noop {} {\bibfield
  {journal} {\bibinfo  {journal} {J. Chem. Phys.}\ }\textbf {\bibinfo {volume}
  {141}},\ \bibinfo {pages} {214704} (\bibinfo {year} {2014})}\BibitemShut
  {NoStop}%
\bibitem [{\citenamefont {Kabir}\ and\ \citenamefont
  {Saha-Dasgupta}(2014)}]{kabirPRB90}%
  \BibitemOpen
  \bibfield  {author} {\bibinfo {author} {\bibfnamefont {M.}~\bibnamefont
  {Kabir}}\ and\ \bibinfo {author} {\bibfnamefont {T.}~\bibnamefont
  {Saha-Dasgupta}},\ }\href@noop {} {\bibfield  {journal} {\bibinfo  {journal}
  {Phys. Rev. B}\ }\textbf {\bibinfo {volume} {90}},\ \bibinfo {pages} {035403}
  (\bibinfo {year} {2014})}\BibitemShut {NoStop}%
\bibitem [{\citenamefont {Ganguly}\ \emph {et~al.}(2017)\citenamefont
  {Ganguly}, \citenamefont {Kabir},\ and\ \citenamefont
  {Saha-Dasgupta}}]{gangulyPRB95}%
  \BibitemOpen
  \bibfield  {author} {\bibinfo {author} {\bibfnamefont {S.}~\bibnamefont
  {Ganguly}}, \bibinfo {author} {\bibfnamefont {M.}~\bibnamefont {Kabir}}, \
  and\ \bibinfo {author} {\bibfnamefont {T.}~\bibnamefont {Saha-Dasgupta}},\
  }\href@noop {} {\bibfield  {journal} {\bibinfo  {journal} {Phys. Rev. B}\
  }\textbf {\bibinfo {volume} {95}},\ \bibinfo {pages} {174419} (\bibinfo
  {year} {2017})}\BibitemShut {NoStop}%
\bibitem [{\citenamefont {Valli}\ \emph {et~al.}(2016)\citenamefont {Valli},
  \citenamefont {Amaricci}, \citenamefont {Toschi}, \citenamefont
  {T.Saha-Dasgupta}, \citenamefont {Held},\ and\ \citenamefont
  {Capone}}]{valliPRB94}%
  \BibitemOpen
  \bibfield  {author} {\bibinfo {author} {\bibfnamefont {A.}~\bibnamefont
  {Valli}}, \bibinfo {author} {\bibfnamefont {A.}~\bibnamefont {Amaricci}},
  \bibinfo {author} {\bibfnamefont {A.}~\bibnamefont {Toschi}}, \bibinfo
  {author} {\bibnamefont {T.Saha-Dasgupta}}, \bibinfo {author} {\bibfnamefont
  {K.}~\bibnamefont {Held}}, \ and\ \bibinfo {author} {\bibfnamefont
  {M.}~\bibnamefont {Capone}},\ }\href@noop {} {\bibfield  {journal} {\bibinfo
  {journal} {Phys. Rev. B}\ }\textbf {\bibinfo {volume} {94}},\ \bibinfo
  {pages} {245146} (\bibinfo {year} {2016})}\BibitemShut {NoStop}%
\bibitem [{\citenamefont {Tao}\ \emph {et~al.}(2011)\citenamefont {Tao},
  \citenamefont {Jiao}, \citenamefont {Yazyev}, \citenamefont {Chen},
  \citenamefont {Feng}, \citenamefont {Zhang}, \citenamefont {Capaz},
  \citenamefont {Tour}, \citenamefont {Zettl}, \citenamefont {Louie},
  \citenamefont {Dai},\ and\ \citenamefont {Crommie}}]{taoNatPhys7}%
  \BibitemOpen
  \bibfield  {author} {\bibinfo {author} {\bibfnamefont {C.}~\bibnamefont
  {Tao}}, \bibinfo {author} {\bibfnamefont {L.}~\bibnamefont {Jiao}}, \bibinfo
  {author} {\bibfnamefont {O.~V.}\ \bibnamefont {Yazyev}}, \bibinfo {author}
  {\bibfnamefont {Y.-C.}\ \bibnamefont {Chen}}, \bibinfo {author}
  {\bibfnamefont {J.}~\bibnamefont {Feng}}, \bibinfo {author} {\bibfnamefont
  {X.}~\bibnamefont {Zhang}}, \bibinfo {author} {\bibfnamefont {R.~B.}\
  \bibnamefont {Capaz}}, \bibinfo {author} {\bibfnamefont {J.~M.}\ \bibnamefont
  {Tour}}, \bibinfo {author} {\bibfnamefont {A.}~\bibnamefont {Zettl}},
  \bibinfo {author} {\bibfnamefont {S.~G.}\ \bibnamefont {Louie}}, \bibinfo
  {author} {\bibfnamefont {H.}~\bibnamefont {Dai}}, \ and\ \bibinfo {author}
  {\bibfnamefont {M.~F.}\ \bibnamefont {Crommie}},\ }\href@noop {} {\bibfield
  {journal} {\bibinfo  {journal} {Nat. Phys.}\ }\textbf {\bibinfo {volume}
  {7}},\ \bibinfo {pages} {616} (\bibinfo {year} {2011})}\BibitemShut {NoStop}%
\bibitem [{\citenamefont {Magda}\ \emph {et~al.}(2014)\citenamefont {Magda},
  \citenamefont {Jin}, \citenamefont {Hagym\'{a}si}, \citenamefont
  {Vancs\'{o}}, \citenamefont {Osv\'{a}th}, \citenamefont {Nemes-Incze},
  \citenamefont {Hwang}, \citenamefont {Bir\'{o}},\ and\ \citenamefont
  {Tapaszt\'{o}}}]{magdaNat514}%
  \BibitemOpen
  \bibfield  {author} {\bibinfo {author} {\bibfnamefont {G.~Z.}\ \bibnamefont
  {Magda}}, \bibinfo {author} {\bibfnamefont {X.}~\bibnamefont {Jin}}, \bibinfo
  {author} {\bibfnamefont {I.}~\bibnamefont {Hagym\'{a}si}}, \bibinfo {author}
  {\bibfnamefont {P.}~\bibnamefont {Vancs\'{o}}}, \bibinfo {author}
  {\bibfnamefont {Z.}~\bibnamefont {Osv\'{a}th}}, \bibinfo {author}
  {\bibfnamefont {P.}~\bibnamefont {Nemes-Incze}}, \bibinfo {author}
  {\bibfnamefont {C.}~\bibnamefont {Hwang}}, \bibinfo {author} {\bibfnamefont
  {L.~P.}\ \bibnamefont {Bir\'{o}}}, \ and\ \bibinfo {author} {\bibfnamefont
  {L.}~\bibnamefont {Tapaszt\'{o}}},\ }\href@noop {} {\bibfield  {journal}
  {\bibinfo  {journal} {Nature}\ }\textbf {\bibinfo {volume} {514}},\ \bibinfo
  {pages} {608} (\bibinfo {year} {2014})}\BibitemShut {NoStop}%
\bibitem [{\citenamefont {Sun}\ \emph {et~al.}(2017)\citenamefont {Sun},
  \citenamefont {Zheng}, \citenamefont {Pan}, \citenamefont {Chen},
  \citenamefont {Zhang}, \citenamefont {Fu}, \citenamefont {Zhang},
  \citenamefont {Tang},\ and\ \citenamefont {Du}}]{sunQM2}%
  \BibitemOpen
  \bibfield  {author} {\bibinfo {author} {\bibfnamefont {Y.}~\bibnamefont
  {Sun}}, \bibinfo {author} {\bibfnamefont {Y.}~\bibnamefont {Zheng}}, \bibinfo
  {author} {\bibfnamefont {H.}~\bibnamefont {Pan}}, \bibinfo {author}
  {\bibfnamefont {J.}~\bibnamefont {Chen}}, \bibinfo {author} {\bibfnamefont
  {W.}~\bibnamefont {Zhang}}, \bibinfo {author} {\bibfnamefont
  {L.}~\bibnamefont {Fu}}, \bibinfo {author} {\bibfnamefont {K.}~\bibnamefont
  {Zhang}}, \bibinfo {author} {\bibfnamefont {N.}~\bibnamefont {Tang}}, \ and\
  \bibinfo {author} {\bibfnamefont {Y.}~\bibnamefont {Du}},\ }\href@noop {}
  {\bibfield  {journal} {\bibinfo  {journal} {npj Quantum Materials}\ }\textbf
  {\bibinfo {volume} {2}},\ \bibinfo {pages} {5} (\bibinfo {year}
  {2017})}\BibitemShut {NoStop}%
\bibitem [{\citenamefont {Yu}\ and\ \citenamefont {Liu}(2016)}]{yuSR6}%
  \BibitemOpen
  \bibfield  {author} {\bibinfo {author} {\bibfnamefont {H.}~\bibnamefont
  {Yu}}\ and\ \bibinfo {author} {\bibfnamefont {J.-F.}\ \bibnamefont {Liu}},\
  }\href@noop {} {\bibfield  {journal} {\bibinfo  {journal} {Sci. Rep.}\
  }\textbf {\bibinfo {volume} {6}},\ \bibinfo {pages} {25361} (\bibinfo {year}
  {2016})}\BibitemShut {NoStop}%
\bibitem [{\citenamefont {Ozaki}\ \emph {et~al.}(2010)\citenamefont {Ozaki},
  \citenamefont {Nishio}, \citenamefont {Weng},\ and\ \citenamefont
  {Kino}}]{ozakiPRB81}%
  \BibitemOpen
  \bibfield  {author} {\bibinfo {author} {\bibfnamefont {T.}~\bibnamefont
  {Ozaki}}, \bibinfo {author} {\bibfnamefont {K.}~\bibnamefont {Nishio}},
  \bibinfo {author} {\bibfnamefont {H.}~\bibnamefont {Weng}}, \ and\ \bibinfo
  {author} {\bibfnamefont {H.}~\bibnamefont {Kino}},\ }\href@noop {} {\bibfield
   {journal} {\bibinfo  {journal} {Phys. Rev. B}\ }\textbf {\bibinfo {volume}
  {81}},\ \bibinfo {pages} {075422} (\bibinfo {year} {2010})}\BibitemShut
  {NoStop}%
\bibitem [{\citenamefont {Cocchi}\ \emph {et~al.}(2010)\citenamefont {Cocchi},
  \citenamefont {Prezzi}, \citenamefont {Calzolari},\ and\ \citenamefont
  {Molinari}}]{cocchiJCP133}%
  \BibitemOpen
  \bibfield  {author} {\bibinfo {author} {\bibfnamefont {C.}~\bibnamefont
  {Cocchi}}, \bibinfo {author} {\bibfnamefont {D.}~\bibnamefont {Prezzi}},
  \bibinfo {author} {\bibfnamefont {A.}~\bibnamefont {Calzolari}}, \ and\
  \bibinfo {author} {\bibfnamefont {E.}~\bibnamefont {Molinari}},\ }\href@noop
  {} {\bibfield  {journal} {\bibinfo  {journal} {J. Chem. Phys.}\ }\textbf
  {\bibinfo {volume} {133}},\ \bibinfo {pages} {124703} (\bibinfo {year}
  {2010})}\BibitemShut {NoStop}%
\bibitem [{\citenamefont {Chen}\ \emph {et~al.}(2013)\citenamefont {Chen},
  \citenamefont {Meng}, \citenamefont {Zhou}, \citenamefont {Wu}, \citenamefont
  {Bie}, \citenamefont {Liao},\ and\ \citenamefont {Yu}}]{chenNatComm4}%
  \BibitemOpen
  \bibfield  {author} {\bibinfo {author} {\bibfnamefont {J.-J.}\ \bibnamefont
  {Chen}}, \bibinfo {author} {\bibfnamefont {J.}~\bibnamefont {Meng}}, \bibinfo
  {author} {\bibfnamefont {Y.-B.}\ \bibnamefont {Zhou}}, \bibinfo {author}
  {\bibfnamefont {H.-C.}\ \bibnamefont {Wu}}, \bibinfo {author} {\bibfnamefont
  {Y.-Q.}\ \bibnamefont {Bie}}, \bibinfo {author} {\bibfnamefont {Z.-M.}\
  \bibnamefont {Liao}}, \ and\ \bibinfo {author} {\bibfnamefont {D.-P.}\
  \bibnamefont {Yu}},\ }\href@noop {} {\bibfield  {journal} {\bibinfo
  {journal} {Nat. Commun.}\ }\textbf {\bibinfo {volume} {4}},\ \bibinfo {pages}
  {1921} (\bibinfo {year} {2013})}\BibitemShut {NoStop}%
\bibitem [{\citenamefont {Lundeberg}\ and\ \citenamefont
  {Folk}(2009)}]{lundebergNatPhys5}%
  \BibitemOpen
  \bibfield  {author} {\bibinfo {author} {\bibfnamefont {M.~B.}\ \bibnamefont
  {Lundeberg}}\ and\ \bibinfo {author} {\bibfnamefont {J.~A.}\ \bibnamefont
  {Folk}},\ }\href@noop {} {\bibfield  {journal} {\bibinfo  {journal} {Nat.
  Phys.}\ }\textbf {\bibinfo {volume} {5}},\ \bibinfo {pages} {894} (\bibinfo
  {year} {2009})}\BibitemShut {NoStop}%
\bibitem [{\citenamefont {Kretinin}\ \emph {et~al.}(2013)\citenamefont
  {Kretinin}, \citenamefont {Yu}, \citenamefont {Jalil}, \citenamefont {Cao},
  \citenamefont {Withers}, \citenamefont {Mishchenko}, \citenamefont
  {Katsnelson}, \citenamefont {Novoselov}, \citenamefont {Geim},\ and\
  \citenamefont {Guinea}}]{kretininPRB88}%
  \BibitemOpen
  \bibfield  {author} {\bibinfo {author} {\bibfnamefont {A.}~\bibnamefont
  {Kretinin}}, \bibinfo {author} {\bibfnamefont {G.~L.}\ \bibnamefont {Yu}},
  \bibinfo {author} {\bibfnamefont {R.}~\bibnamefont {Jalil}}, \bibinfo
  {author} {\bibfnamefont {Y.}~\bibnamefont {Cao}}, \bibinfo {author}
  {\bibfnamefont {F.}~\bibnamefont {Withers}}, \bibinfo {author} {\bibfnamefont
  {A.}~\bibnamefont {Mishchenko}}, \bibinfo {author} {\bibfnamefont {M.~I.}\
  \bibnamefont {Katsnelson}}, \bibinfo {author} {\bibfnamefont {K.~S.}\
  \bibnamefont {Novoselov}}, \bibinfo {author} {\bibfnamefont {A.~K.}\
  \bibnamefont {Geim}}, \ and\ \bibinfo {author} {\bibfnamefont
  {F.}~\bibnamefont {Guinea}},\ }\href@noop {} {\bibfield  {journal} {\bibinfo
  {journal} {Phys. Rev. B}\ }\textbf {\bibinfo {volume} {88}},\ \bibinfo
  {pages} {165427} (\bibinfo {year} {2013})}\BibitemShut {NoStop}%
\bibitem [{\citenamefont {Landauer}(1957)}]{landauerJRD1}%
  \BibitemOpen
  \bibfield  {author} {\bibinfo {author} {\bibfnamefont {R.}~\bibnamefont
  {Landauer}},\ }\href@noop {} {\bibfield  {journal} {\bibinfo  {journal} {IBM
  J. Res. Dev.}\ }\textbf {\bibinfo {volume} {1}},\ \bibinfo {pages} {223}
  (\bibinfo {year} {1957})}\BibitemShut {NoStop}%
\bibitem [{\citenamefont {Meir}\ and\ \citenamefont
  {Wingreen}(1992)}]{meirPRL68}%
  \BibitemOpen
  \bibfield  {author} {\bibinfo {author} {\bibfnamefont {Y.}~\bibnamefont
  {Meir}}\ and\ \bibinfo {author} {\bibfnamefont {N.~S.}\ \bibnamefont
  {Wingreen}},\ }\href@noop {} {\bibfield  {journal} {\bibinfo  {journal}
  {Phys. Rev. Lett.}\ }\textbf {\bibinfo {volume} {68}},\ \bibinfo {pages}
  {2512} (\bibinfo {year} {1992})}\BibitemShut {NoStop}%
\bibitem [{\citenamefont {Datta}(1995)}]{dattaETMS}%
  \BibitemOpen
  \bibfield  {author} {\bibinfo {author} {\bibfnamefont {S.}~\bibnamefont
  {Datta}},\ }\href {\doibase 10.1017/CBO9780511805776} {\emph {\bibinfo
  {title} {Electronic Transport in Mesoscopic Systems}}},\ Cambridge Studies in
  Semiconductor Physics and Microelectronic Engineering\ (\bibinfo  {publisher}
  {Cambridge University Press},\ \bibinfo {year} {1995})\BibitemShut {NoStop}%
\bibitem [{\citenamefont {Georges}\ \emph {et~al.}(1996)\citenamefont
  {Georges}, \citenamefont {Kotliar}, \citenamefont {Krauth},\ and\
  \citenamefont {Rozenberg}}]{georgesRMP69}%
  \BibitemOpen
  \bibfield  {author} {\bibinfo {author} {\bibfnamefont {A.}~\bibnamefont
  {Georges}}, \bibinfo {author} {\bibfnamefont {G.}~\bibnamefont {Kotliar}},
  \bibinfo {author} {\bibfnamefont {W.}~\bibnamefont {Krauth}}, \ and\ \bibinfo
  {author} {\bibfnamefont {M.}~\bibnamefont {Rozenberg}},\ }\href@noop {}
  {\bibfield  {journal} {\bibinfo  {journal} {Rev. Mod. Phys.}\ }\textbf
  {\bibinfo {volume} {68}},\ \bibinfo {pages} {13} (\bibinfo {year}
  {1996})}\BibitemShut {NoStop}%
\bibitem [{\citenamefont {Snoek}\ \emph {et~al.}(2008)\citenamefont {Snoek},
  \citenamefont {Titvinidze}, \citenamefont {T{\"o}ke}, \citenamefont
  {Byczuk},\ and\ \citenamefont {Hofstetter}}]{snoekNJP10}%
  \BibitemOpen
  \bibfield  {author} {\bibinfo {author} {\bibfnamefont {M.}~\bibnamefont
  {Snoek}}, \bibinfo {author} {\bibfnamefont {I.}~\bibnamefont {Titvinidze}},
  \bibinfo {author} {\bibfnamefont {C.}~\bibnamefont {T{\"o}ke}}, \bibinfo
  {author} {\bibfnamefont {K.}~\bibnamefont {Byczuk}}, \ and\ \bibinfo {author}
  {\bibfnamefont {W.}~\bibnamefont {Hofstetter}},\ }\href@noop {} {\bibfield
  {journal} {\bibinfo  {journal} {New J. Phys.}\ }\textbf {\bibinfo {volume}
  {10}},\ \bibinfo {pages} {093008} (\bibinfo {year} {2008})}\BibitemShut
  {NoStop}%
\bibitem [{\citenamefont {Valli}\ \emph {et~al.}(2010)\citenamefont {Valli},
  \citenamefont {Sangiovanni}, \citenamefont {Gunnarsson}, \citenamefont
  {Toschi},\ and\ \citenamefont {Held}}]{valliPRL104}%
  \BibitemOpen
  \bibfield  {author} {\bibinfo {author} {\bibfnamefont {A.}~\bibnamefont
  {Valli}}, \bibinfo {author} {\bibfnamefont {G.}~\bibnamefont {Sangiovanni}},
  \bibinfo {author} {\bibfnamefont {O.}~\bibnamefont {Gunnarsson}}, \bibinfo
  {author} {\bibfnamefont {A.}~\bibnamefont {Toschi}}, \ and\ \bibinfo {author}
  {\bibfnamefont {K.}~\bibnamefont {Held}},\ }\href@noop {} {\bibfield
  {journal} {\bibinfo  {journal} {Phys. Rev. Lett.}\ }\textbf {\bibinfo
  {volume} {104}},\ \bibinfo {pages} {246402} (\bibinfo {year}
  {2010})}\BibitemShut {NoStop}%
\bibitem [{\citenamefont {Valli}\ \emph {et~al.}(2012)\citenamefont {Valli},
  \citenamefont {Sangiovanni}, \citenamefont {Toschi},\ and\ \citenamefont
  {Held}}]{valliPRB86}%
  \BibitemOpen
  \bibfield  {author} {\bibinfo {author} {\bibfnamefont {A.}~\bibnamefont
  {Valli}}, \bibinfo {author} {\bibfnamefont {G.}~\bibnamefont {Sangiovanni}},
  \bibinfo {author} {\bibfnamefont {A.}~\bibnamefont {Toschi}}, \ and\ \bibinfo
  {author} {\bibfnamefont {K.}~\bibnamefont {Held}},\ }\href@noop {} {\bibfield
   {journal} {\bibinfo  {journal} {Phys. Rev. B}\ } (\bibinfo {year}
  {2012})}\BibitemShut {NoStop}%
\bibitem [{\citenamefont {Jacob}\ \emph {et~al.}(2010)\citenamefont {Jacob},
  \citenamefont {Haule},\ and\ \citenamefont {Kotliar}}]{jacobPRB82}%
  \BibitemOpen
  \bibfield  {author} {\bibinfo {author} {\bibfnamefont {D.}~\bibnamefont
  {Jacob}}, \bibinfo {author} {\bibfnamefont {K.}~\bibnamefont {Haule}}, \ and\
  \bibinfo {author} {\bibfnamefont {G.}~\bibnamefont {Kotliar}},\ }\href@noop
  {} {\bibfield  {journal} {\bibinfo  {journal} {Phys. Rev. B}\ }\textbf
  {\bibinfo {volume} {82}},\ \bibinfo {pages} {195115} (\bibinfo {year}
  {2010})}\BibitemShut {NoStop}%
\bibitem [{\citenamefont {Mazza}\ \emph {et~al.}(2016)\citenamefont {Mazza},
  \citenamefont {Amaricci}, \citenamefont {Capone},\ and\ \citenamefont
  {Fabrizio}}]{mazzaPRL117}%
  \BibitemOpen
  \bibfield  {author} {\bibinfo {author} {\bibfnamefont {G.}~\bibnamefont
  {Mazza}}, \bibinfo {author} {\bibfnamefont {A.}~\bibnamefont {Amaricci}},
  \bibinfo {author} {\bibfnamefont {M.}~\bibnamefont {Capone}}, \ and\ \bibinfo
  {author} {\bibfnamefont {M.}~\bibnamefont {Fabrizio}},\ }\href@noop {}
  {\bibfield  {journal} {\bibinfo  {journal} {Phys. Rev. Lett.}\ }\textbf
  {\bibinfo {volume} {117}},\ \bibinfo {pages} {176401} (\bibinfo {year}
  {2016})}\BibitemShut {NoStop}%
\bibitem [{\citenamefont {Sch\"{u}ler}\ \emph {et~al.}(2017)\citenamefont
  {Sch\"{u}ler}, \citenamefont {Barthel}, \citenamefont {Wehling},
  \citenamefont {Karolak}, \citenamefont {Valli},\ and\ \citenamefont
  {Sangiovanni}}]{schuelerEPJST226}%
  \BibitemOpen
  \bibfield  {author} {\bibinfo {author} {\bibfnamefont {M.}~\bibnamefont
  {Sch\"{u}ler}}, \bibinfo {author} {\bibfnamefont {S.}~\bibnamefont
  {Barthel}}, \bibinfo {author} {\bibfnamefont {T.}~\bibnamefont {Wehling}},
  \bibinfo {author} {\bibfnamefont {M.}~\bibnamefont {Karolak}}, \bibinfo
  {author} {\bibfnamefont {A.}~\bibnamefont {Valli}}, \ and\ \bibinfo {author}
  {\bibfnamefont {G.}~\bibnamefont {Sangiovanni}},\ }\href@noop {} {\bibfield
  {journal} {\bibinfo  {journal} {Eur. Phys. J. Special Topics}\ }\textbf
  {\bibinfo {volume} {226}},\ \bibinfo {pages} {2615} (\bibinfo {year}
  {2017})}\BibitemShut {NoStop}%
\bibitem [{\citenamefont {Amaricci}\ \emph {et~al.}(2014)\citenamefont
  {Amaricci}, \citenamefont {Privitera},\ and\ \citenamefont
  {Capone}}]{amaricciPRA89}%
  \BibitemOpen
  \bibfield  {author} {\bibinfo {author} {\bibfnamefont {A.}~\bibnamefont
  {Amaricci}}, \bibinfo {author} {\bibfnamefont {A.}~\bibnamefont {Privitera}},
  \ and\ \bibinfo {author} {\bibfnamefont {M.}~\bibnamefont {Capone}},\
  }\href@noop {} {\bibfield  {journal} {\bibinfo  {journal} {Phys. Rev. A}\
  }\textbf {\bibinfo {volume} {89}},\ \bibinfo {pages} {053604} (\bibinfo
  {year} {2014})}\BibitemShut {NoStop}%
\bibitem [{\citenamefont {Solomon}\ \emph
  {et~al.}(2008{\natexlab{b}})\citenamefont {Solomon}, \citenamefont {Andrews},
  \citenamefont {Van~Duyne},\ and\ \citenamefont {Ratner}}]{solomonJACS130}%
  \BibitemOpen
  \bibfield  {author} {\bibinfo {author} {\bibfnamefont {G.~C.}\ \bibnamefont
  {Solomon}}, \bibinfo {author} {\bibfnamefont {D.~Q.}\ \bibnamefont
  {Andrews}}, \bibinfo {author} {\bibfnamefont {R.~P.}\ \bibnamefont
  {Van~Duyne}}, \ and\ \bibinfo {author} {\bibfnamefont {M.~A.}\ \bibnamefont
  {Ratner}},\ }\href@noop {} {\bibfield  {journal} {\bibinfo  {journal} {J. Am.
  Chem. Soc.}\ }\textbf {\bibinfo {volume} {130}},\ \bibinfo {pages} {7788}
  (\bibinfo {year} {2008}{\natexlab{b}})}\BibitemShut {NoStop}%
\bibitem [{\citenamefont {Feldner}\ \emph {et~al.}(2010)\citenamefont
  {Feldner}, \citenamefont {Meng}, \citenamefont {Honecker}, \citenamefont
  {Cabra}, \citenamefont {Wessel},\ and\ \citenamefont
  {Assaad}}]{feldnerPRB81}%
  \BibitemOpen
  \bibfield  {author} {\bibinfo {author} {\bibfnamefont {H.}~\bibnamefont
  {Feldner}}, \bibinfo {author} {\bibfnamefont {Z.~Y.}\ \bibnamefont {Meng}},
  \bibinfo {author} {\bibfnamefont {A.}~\bibnamefont {Honecker}}, \bibinfo
  {author} {\bibfnamefont {D.}~\bibnamefont {Cabra}}, \bibinfo {author}
  {\bibfnamefont {S.}~\bibnamefont {Wessel}}, \ and\ \bibinfo {author}
  {\bibfnamefont {F.~F.}\ \bibnamefont {Assaad}},\ }\href@noop {} {\bibfield
  {journal} {\bibinfo  {journal} {Phys. Rev. B}\ }\textbf {\bibinfo {volume}
  {81}},\ \bibinfo {pages} {115416} (\bibinfo {year} {2010})}\BibitemShut
  {NoStop}%
\bibitem [{\citenamefont {Dean}\ \emph {et~al.}(2010)\citenamefont {Dean},
  \citenamefont {Young}, \citenamefont {Meric}, \citenamefont {Lee},
  \citenamefont {Wang}, \citenamefont {Sorgenfrei}, \citenamefont {Watanabe},
  \citenamefont {Taniguchi}, \citenamefont {Kim}, \citenamefont {Shepard},\
  and\ \citenamefont {Hone}}]{deanNatNanotech5}%
  \BibitemOpen
  \bibfield  {author} {\bibinfo {author} {\bibfnamefont {C.~R.}\ \bibnamefont
  {Dean}}, \bibinfo {author} {\bibfnamefont {A.~F.}\ \bibnamefont {Young}},
  \bibinfo {author} {\bibfnamefont {I.}~\bibnamefont {Meric}}, \bibinfo
  {author} {\bibfnamefont {C.}~\bibnamefont {Lee}}, \bibinfo {author}
  {\bibfnamefont {L.}~\bibnamefont {Wang}}, \bibinfo {author} {\bibfnamefont
  {S.}~\bibnamefont {Sorgenfrei}}, \bibinfo {author} {\bibfnamefont
  {K.}~\bibnamefont {Watanabe}}, \bibinfo {author} {\bibfnamefont
  {T.}~\bibnamefont {Taniguchi}}, \bibinfo {author} {\bibfnamefont
  {P.}~\bibnamefont {Kim}}, \bibinfo {author} {\bibfnamefont {K.~L.}\
  \bibnamefont {Shepard}}, \ and\ \bibinfo {author} {\bibfnamefont
  {J.}~\bibnamefont {Hone}},\ }\href@noop {} {\bibfield  {journal} {\bibinfo
  {journal} {Nat. Nanotech.}\ }\textbf {\bibinfo {volume} {5}},\ \bibinfo
  {pages} {722} (\bibinfo {year} {2010})}\BibitemShut {NoStop}%
\bibitem [{\citenamefont {Giovannetti}\ \emph {et~al.}(2007)\citenamefont
  {Giovannetti}, \citenamefont {Khomyakov}, \citenamefont {Brocks},
  \citenamefont {J.Kelly},\ and\ \citenamefont {van~den
  Brink}}]{giovannettiPRB76}%
  \BibitemOpen
  \bibfield  {author} {\bibinfo {author} {\bibfnamefont {G.}~\bibnamefont
  {Giovannetti}}, \bibinfo {author} {\bibfnamefont {P.~A.}\ \bibnamefont
  {Khomyakov}}, \bibinfo {author} {\bibfnamefont {G.}~\bibnamefont {Brocks}},
  \bibinfo {author} {\bibfnamefont {P.}~\bibnamefont {J.Kelly}}, \ and\
  \bibinfo {author} {\bibfnamefont {J.}~\bibnamefont {van~den Brink}},\
  }\href@noop {} {\bibfield  {journal} {\bibinfo  {journal} {Phys. Rev. B}\
  }\textbf {\bibinfo {volume} {76}},\ \bibinfo {pages} {073103} (\bibinfo
  {year} {2007})}\BibitemShut {NoStop}%
\bibitem [{\citenamefont {Skomski}\ \emph {et~al.}(2014)\citenamefont
  {Skomski}, \citenamefont {Dowben}, \citenamefont {Driver},\ and\
  \citenamefont {Kelber}}]{skomskiMH563}%
  \BibitemOpen
  \bibfield  {author} {\bibinfo {author} {\bibfnamefont {R.}~\bibnamefont
  {Skomski}}, \bibinfo {author} {\bibfnamefont {P.~A.}\ \bibnamefont {Dowben}},
  \bibinfo {author} {\bibfnamefont {M.~S.}\ \bibnamefont {Driver}}, \ and\
  \bibinfo {author} {\bibfnamefont {J.~A.}\ \bibnamefont {Kelber}},\
  }\href@noop {} {\bibfield  {journal} {\bibinfo  {journal} {Mater. Horiz.}\
  }\textbf {\bibinfo {volume} {1}},\ \bibinfo {pages} {563} (\bibinfo {year}
  {2014})}\BibitemShut {NoStop}%
\bibitem [{\citenamefont {Slawi\'{n}ska}\ \emph {et~al.}(2010)\citenamefont
  {Slawi\'{n}ska}, \citenamefont {Zasada},\ and\ \citenamefont
  {Klusek}}]{slawinskaPRB81}%
  \BibitemOpen
  \bibfield  {author} {\bibinfo {author} {\bibfnamefont {J.}~\bibnamefont
  {Slawi\'{n}ska}}, \bibinfo {author} {\bibfnamefont {I.}~\bibnamefont
  {Zasada}}, \ and\ \bibinfo {author} {\bibfnamefont {Z.}~\bibnamefont
  {Klusek}},\ }\href@noop {} {\bibfield  {journal} {\bibinfo  {journal} {Phys.
  Rev. B}\ }\textbf {\bibinfo {volume} {81}},\ \bibinfo {pages} {155433}
  (\bibinfo {year} {2010})}\BibitemShut {NoStop}%
\bibitem [{\citenamefont {Guan}\ \emph {et~al.}(2015)\citenamefont {Guan},
  \citenamefont {Li},\ and\ \citenamefont {Duan}}]{guanAPA123}%
  \BibitemOpen
  \bibfield  {author} {\bibinfo {author} {\bibfnamefont {Z.}~\bibnamefont
  {Guan}}, \bibinfo {author} {\bibfnamefont {J.}~\bibnamefont {Li}}, \ and\
  \bibinfo {author} {\bibfnamefont {W.}~\bibnamefont {Duan}},\ }\href@noop {}
  {\bibfield  {journal} {\bibinfo  {journal} {Appl. Phys. A}\ }\textbf
  {\bibinfo {volume} {123}},\ \bibinfo {pages} {678} (\bibinfo {year}
  {2015})}\BibitemShut {NoStop}%
\bibitem [{\citenamefont {Zhang}\ \emph {et~al.}(2009)\citenamefont {Zhang},
  \citenamefont {Tang}, \citenamefont {Girit}, \citenamefont {Hao},
  \citenamefont {Martin}, \citenamefont {Zettl}, \citenamefont {Crommie},
  \citenamefont {Shen},\ and\ \citenamefont {Wang}}]{zhangNature459}%
  \BibitemOpen
  \bibfield  {author} {\bibinfo {author} {\bibfnamefont {Y.}~\bibnamefont
  {Zhang}}, \bibinfo {author} {\bibfnamefont {T.-T.}\ \bibnamefont {Tang}},
  \bibinfo {author} {\bibfnamefont {C.}~\bibnamefont {Girit}}, \bibinfo
  {author} {\bibfnamefont {Z.}~\bibnamefont {Hao}}, \bibinfo {author}
  {\bibfnamefont {M.~C.}\ \bibnamefont {Martin}}, \bibinfo {author}
  {\bibfnamefont {A.}~\bibnamefont {Zettl}}, \bibinfo {author} {\bibfnamefont
  {M.~F.}\ \bibnamefont {Crommie}}, \bibinfo {author} {\bibfnamefont {Y.~R.}\
  \bibnamefont {Shen}}, \ and\ \bibinfo {author} {\bibfnamefont
  {F.}~\bibnamefont {Wang}},\ }\href@noop {} {\bibfield  {journal} {\bibinfo
  {journal} {Nature}\ }\textbf {\bibinfo {volume} {459}},\ \bibinfo {pages}
  {820} (\bibinfo {year} {2009})}\BibitemShut {NoStop}%
\bibitem [{\citenamefont {Neto}\ \emph {et~al.}(81)\citenamefont {Neto},
  \citenamefont {Guinea}, \citenamefont {Peres}, \citenamefont {Novoselov},\
  and\ \citenamefont {Geim}}]{castronetoRMP81}%
  \BibitemOpen
  \bibfield  {author} {\bibinfo {author} {\bibfnamefont {A.~H.~C.}\
  \bibnamefont {Neto}}, \bibinfo {author} {\bibfnamefont {F.}~\bibnamefont
  {Guinea}}, \bibinfo {author} {\bibfnamefont {N.~M.~R.}\ \bibnamefont
  {Peres}}, \bibinfo {author} {\bibfnamefont {K.~S.}\ \bibnamefont
  {Novoselov}}, \ and\ \bibinfo {author} {\bibfnamefont {A.~K.}\ \bibnamefont
  {Geim}},\ }\href@noop {} {\bibfield  {journal} {\bibinfo  {journal} {Rev.
  Mod. Phys.}\ }\textbf {\bibinfo {volume} {2009}},\ \bibinfo {pages} {109}
  (\bibinfo {year} {81})}\BibitemShut {NoStop}%
\end{thebibliography}

\begin{thebibliography}{10}%
\makeatletter
\providecommand \@ifxundefined [1]{%
 \@ifx{#1\undefined}
}%
\providecommand \@ifnum [1]{%
 \ifnum #1\expandafter \@firstoftwo
 \else \expandafter \@secondoftwo
 \fi
}%
\providecommand \@ifx [1]{%
 \ifx #1\expandafter \@firstoftwo
 \else \expandafter \@secondoftwo
 \fi
}%
\providecommand \natexlab [1]{#1}%
\providecommand \enquote  [1]{``#1''}%
\providecommand \bibnamefont  [1]{#1}%
\providecommand \bibfnamefont [1]{#1}%
\providecommand \citenamefont [1]{#1}%
\providecommand \href@noop [0]{\@secondoftwo}%
\providecommand \href [0]{\begingroup \@sanitize@url \@href}%
\providecommand \@href[1]{\@@startlink{#1}\@@href}%
\providecommand \@@href[1]{\endgroup#1\@@endlink}%
\providecommand \@sanitize@url [0]{\catcode `\\12\catcode `\$12\catcode
  `\&12\catcode `\#12\catcode `\^12\catcode `\_12\catcode `\%12\relax}%
\providecommand \@@startlink[1]{}%
\providecommand \@@endlink[0]{}%
\providecommand \url  [0]{\begingroup\@sanitize@url \@url }%
\providecommand \@url [1]{\endgroup\@href {#1}{\urlprefix }}%
\providecommand \urlprefix  [0]{URL }%
\providecommand \Eprint [0]{\href }%
\providecommand \doibase [0]{http://dx.doi.org/}%
\providecommand \selectlanguage [0]{\@gobble}%
\providecommand \bibinfo  [0]{\@secondoftwo}%
\providecommand \bibfield  [0]{\@secondoftwo}%
\providecommand \translation [1]{[#1]}%
\providecommand \BibitemOpen [0]{}%
\providecommand \bibitemStop [0]{}%
\providecommand \bibitemNoStop [0]{.\EOS\space}%
\providecommand \EOS [0]{\spacefactor3000\relax}%
\providecommand \BibitemShut  [1]{\csname bibitem#1\endcsname}%
\let\auto@bib@innerbib\@empty
\bibitem [{\citenamefont {Parr}\ \emph {et~al.}(1950)\citenamefont {Parr},
  \citenamefont {Craig},\ and\ \citenamefont {Ross}}]{parrJCP18}%
  \BibitemOpen
  \bibfield  {author} {\bibinfo {author} {\bibfnamefont {R.~G.}\ \bibnamefont
  {Parr}}, \bibinfo {author} {\bibfnamefont {D.~P.}\ \bibnamefont {Craig}}, \
  and\ \bibinfo {author} {\bibfnamefont {I.~G.}\ \bibnamefont {Ross}},\
  }\href@noop {} {\bibfield  {journal} {\bibinfo  {journal} {J. Chem Phys.}\
  }\textbf {\bibinfo {volume} {18}},\ \bibinfo {pages} {1561} (\bibinfo {year}
  {1950})}\BibitemShut {NoStop}%
\bibitem [{\citenamefont {Neto}\ \emph {et~al.}(2009)\citenamefont {Neto},
  \citenamefont {Guinea}, \citenamefont {Peres}, \citenamefont {Novoselov},\
  and\ \citenamefont {Geim}}]{castronetoRMP81}%
  \BibitemOpen
  \bibfield  {author} {\bibinfo {author} {\bibfnamefont {A.~H.~C.}\
  \bibnamefont {Neto}}, \bibinfo {author} {\bibfnamefont {F.}~\bibnamefont
  {Guinea}}, \bibinfo {author} {\bibfnamefont {N.~M.~R.}\ \bibnamefont
  {Peres}}, \bibinfo {author} {\bibfnamefont {K.~S.}\ \bibnamefont
  {Novoselov}}, \ and\ \bibinfo {author} {\bibfnamefont {A.~K.}\ \bibnamefont
  {Geim}},\ }\href@noop {} {\bibfield  {journal} {\bibinfo  {journal} {Rev.
  Mod. Phys.}\ }\textbf {\bibinfo {volume} {81}},\ \bibinfo {pages} {109}
  (\bibinfo {year} {2009})}\BibitemShut {NoStop}%
\bibitem [{\citenamefont {Ritter}\ and\ \citenamefont
  {Lyding}(2009)}]{ritterNM8}%
  \BibitemOpen
  \bibfield  {author} {\bibinfo {author} {\bibfnamefont {K.~A.}\ \bibnamefont
  {Ritter}}\ and\ \bibinfo {author} {\bibfnamefont {J.~W.}\ \bibnamefont
  {Lyding}},\ }\href@noop {} {\bibfield  {journal} {\bibinfo  {journal} {Nat.
  Mat.}\ }\textbf {\bibinfo {volume} {8}},\ \bibinfo {pages} {235} (\bibinfo
  {year} {2009})}\BibitemShut {NoStop}%
\bibitem [{\citenamefont {Hu}\ \emph {et~al.}(2014)\citenamefont {Hu},
  \citenamefont {Lin}, \citenamefont {Yang},\ and\ \citenamefont
  {Yang}}]{huJCP141}%
  \BibitemOpen
  \bibfield  {author} {\bibinfo {author} {\bibfnamefont {W.}~\bibnamefont
  {Hu}}, \bibinfo {author} {\bibfnamefont {L.}~\bibnamefont {Lin}}, \bibinfo
  {author} {\bibfnamefont {C.}~\bibnamefont {Yang}}, \ and\ \bibinfo {author}
  {\bibfnamefont {J.}~\bibnamefont {Yang}},\ }\href@noop {} {\bibfield
  {journal} {\bibinfo  {journal} {J. Chem. Phys.}\ }\textbf {\bibinfo {volume}
  {141}},\ \bibinfo {pages} {214704} (\bibinfo {year} {2014})}\BibitemShut
  {NoStop}%
\bibitem [{\citenamefont {Fern\'{a}ndez-Rossier}\ and\ \citenamefont
  {Palacios}(2007)}]{rossierPRL99}%
  \BibitemOpen
  \bibfield  {author} {\bibinfo {author} {\bibfnamefont {J.}~\bibnamefont
  {Fern\'{a}ndez-Rossier}}\ and\ \bibinfo {author} {\bibfnamefont {J.~J.}\
  \bibnamefont {Palacios}},\ }\href@noop {} {\bibfield  {journal} {\bibinfo
  {journal} {Phys. Rev. Lett.}\ }\textbf {\bibinfo {volume} {99}},\ \bibinfo
  {pages} {177204} (\bibinfo {year} {2007})}\BibitemShut {NoStop}%
\bibitem [{\citenamefont {Gunlycke}\ \emph {et~al.}(2007)\citenamefont
  {Gunlycke}, \citenamefont {Areshkin}, \citenamefont {Li}, \citenamefont
  {Mintmire},\ and\ \citenamefont {White}}]{gunlyckeNL7}%
  \BibitemOpen
  \bibfield  {author} {\bibinfo {author} {\bibfnamefont {D.}~\bibnamefont
  {Gunlycke}}, \bibinfo {author} {\bibfnamefont {D.~A.}\ \bibnamefont
  {Areshkin}}, \bibinfo {author} {\bibfnamefont {J.}~\bibnamefont {Li}},
  \bibinfo {author} {\bibfnamefont {J.~W.}\ \bibnamefont {Mintmire}}, \ and\
  \bibinfo {author} {\bibfnamefont {C.~T.}\ \bibnamefont {White}},\ }\href@noop
  {} {\bibfield  {journal} {\bibinfo  {journal} {Nano Lett.}\ }\textbf
  {\bibinfo {volume} {7}},\ \bibinfo {pages} {3608} (\bibinfo {year}
  {2007})}\BibitemShut {NoStop}%
\bibitem [{\citenamefont {Valli}\ \emph {et~al.}(2016)\citenamefont {Valli},
  \citenamefont {Amaricci}, \citenamefont {Toschi}, \citenamefont
  {Saha-Dasgupta}, \citenamefont {Held},\ and\ \citenamefont
  {Capone}}]{valliPRB94}%
  \BibitemOpen
  \bibfield  {author} {\bibinfo {author} {\bibfnamefont {A.}~\bibnamefont
  {Valli}}, \bibinfo {author} {\bibfnamefont {A.}~\bibnamefont {Amaricci}},
  \bibinfo {author} {\bibfnamefont {A.}~\bibnamefont {Toschi}}, \bibinfo
  {author} {\bibfnamefont {T.}~\bibnamefont {Saha-Dasgupta}}, \bibinfo {author}
  {\bibfnamefont {K.}~\bibnamefont {Held}}, \ and\ \bibinfo {author}
  {\bibfnamefont {M.}~\bibnamefont {Capone}},\ }\href@noop {} {\bibfield
  {journal} {\bibinfo  {journal} {Phys. Rev. B}\ }\textbf {\bibinfo {volume}
  {94}},\ \bibinfo {pages} {245146} (\bibinfo {year} {2016})}\BibitemShut
  {NoStop}%
\bibitem [{\citenamefont {Verzijl}\ \emph {et~al.}(2013)\citenamefont
  {Verzijl}, \citenamefont {Seldenthuis},\ and\ \citenamefont
  {Thijssen}}]{verzijlJCP138}%
  \BibitemOpen
  \bibfield  {author} {\bibinfo {author} {\bibfnamefont {C.~J.~O.}\
  \bibnamefont {Verzijl}}, \bibinfo {author} {\bibfnamefont {J.~S.}\
  \bibnamefont {Seldenthuis}}, \ and\ \bibinfo {author} {\bibfnamefont {J.~M.}\
  \bibnamefont {Thijssen}},\ }\href@noop {} {\bibfield  {journal} {\bibinfo
  {journal} {J. Chem. Phys.}\ }\textbf {\bibinfo {volume} {138}},\ \bibinfo
  {pages} {094102} (\bibinfo {year} {2013})}\BibitemShut {NoStop}%
\bibitem [{\citenamefont {Frisenda}\ \emph {et~al.}(2016)\citenamefont
  {Frisenda}, \citenamefont {Janssen}, \citenamefont {Grozema}, \citenamefont
  {der Zant},\ and\ \citenamefont {Renaud}}]{frisendaNatChem8}%
  \BibitemOpen
  \bibfield  {author} {\bibinfo {author} {\bibfnamefont {R.}~\bibnamefont
  {Frisenda}}, \bibinfo {author} {\bibfnamefont {V.~A.~E.~C.}\ \bibnamefont
  {Janssen}}, \bibinfo {author} {\bibfnamefont {F.~C.}\ \bibnamefont
  {Grozema}}, \bibinfo {author} {\bibfnamefont {H.~S.}\ \bibnamefont {der
  Zant}}, \ and\ \bibinfo {author} {\bibfnamefont {N.}~\bibnamefont {Renaud}},\
  }\href@noop {} {\bibfield  {journal} {\bibinfo  {journal} {Nat. Chem.}\
  }\textbf {\bibinfo {volume} {8}},\ \bibinfo {pages} {1099} (\bibinfo {year}
  {2016})}\BibitemShut {NoStop}%
\bibitem [{\citenamefont {Skomski}\ \emph {et~al.}(2014)\citenamefont
  {Skomski}, \citenamefont {Dowben}, \citenamefont {Driver},\ and\
  \citenamefont {Kelber}}]{skomskiMH563}%
  \BibitemOpen
  \bibfield  {author} {\bibinfo {author} {\bibfnamefont {R.}~\bibnamefont
  {Skomski}}, \bibinfo {author} {\bibfnamefont {P.~A.}\ \bibnamefont {Dowben}},
  \bibinfo {author} {\bibfnamefont {M.~S.}\ \bibnamefont {Driver}}, \ and\
  \bibinfo {author} {\bibfnamefont {J.~A.}\ \bibnamefont {Kelber}},\
  }\href@noop {} {\bibfield  {journal} {\bibinfo  {journal} {Mater. Horiz.}\
  }\textbf {\bibinfo {volume} {1}},\ \bibinfo {pages} {563} (\bibinfo {year}
  {2014})}\BibitemShut {NoStop}%
\end{thebibliography}
%


\pagebreak
\onecolumngrid

\thispagestyle{empty}

\begin{center}
  \textbf{\large Table of Content graphics: \\Quantum interference assisted spin filtering in graphene nanoflakes}\\[.2cm]
  A.~Valli,$^{1,*}$ A.~Amaricci,$^{1}$ V.~Brosco,$^1$ and M.~Capone$^1$\\[.1cm]
  {\itshape ${}^1$Scuola Internazionale Superiore di Studi Avanzati (SISSA) \\
CNR-IOM DEMOCRITOS, Istituto Officina dei Materiali, Consiglio Nazionale delle Ricerche, \\
Via Bonomea 265, 34136 Trieste, Italy}\\
\end{center}

\begin{figure}[h]
\includegraphics[width=0.8\textwidth, angle=0]{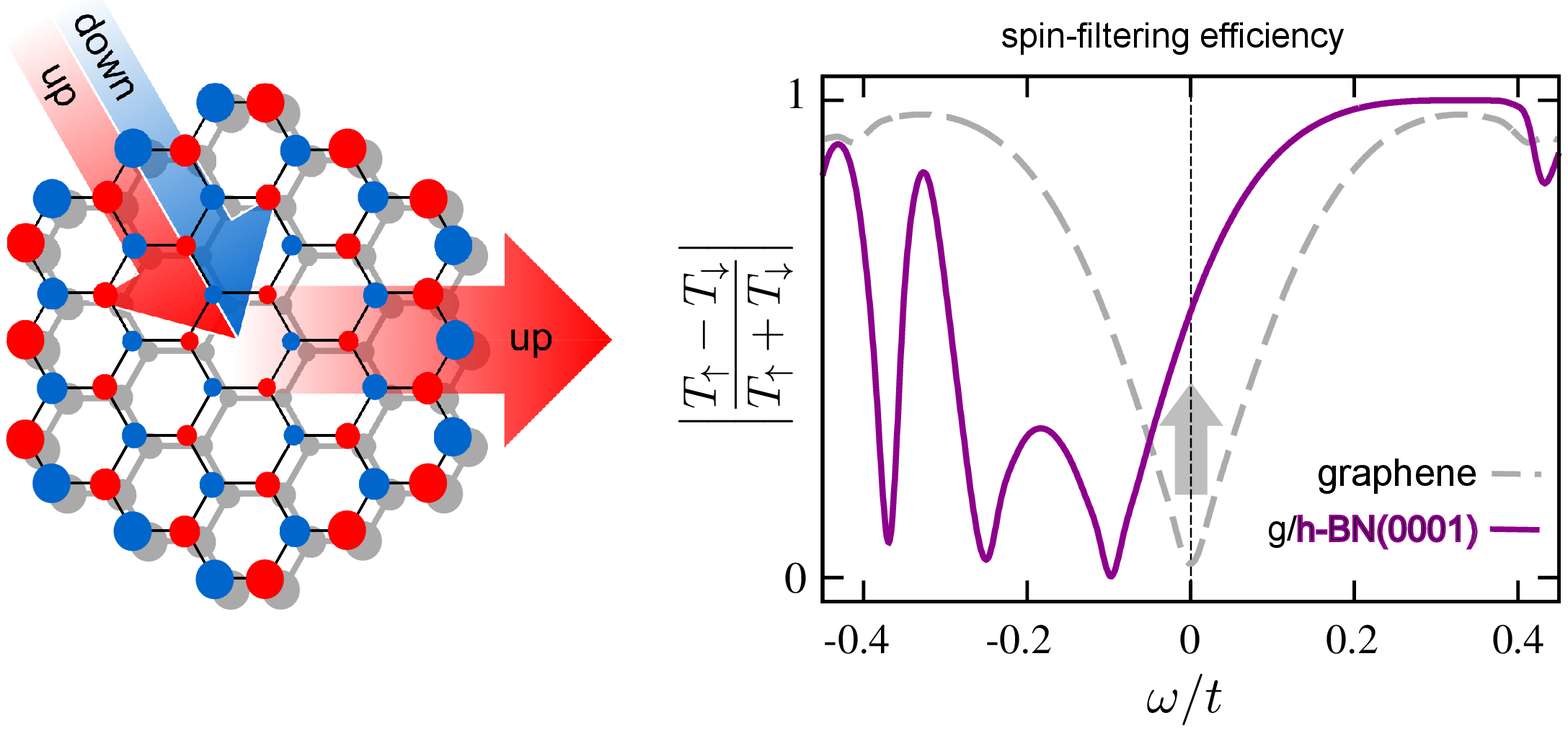}
\label{fig:TOC}
\end{figure}

\end{document}